\documentclass[preprintnumbers,10pt,a4paper,superscriptaddress,aps,pre, twocolumn]{revtex4-2}

\usepackage{amsfonts}
\usepackage{amsmath}
\usepackage{amssymb}
\usepackage{xcolor}
\usepackage{dsfont}
\usepackage{stmaryrd}
\usepackage{verbatim}
\usepackage{bbold}
\usepackage{bm}

\usepackage{graphicx}
\graphicspath{ {./Figures/} }

\usepackage[utf8]{inputenc}
\usepackage{import}
\usepackage{mathtools}
\usepackage{tcolorbox}

\usepackage{xr} 

\usepackage[hidelinks,pagebackref]{hyperref} 
\usepackage[capitalise]{cleveref} 

\crefname{pluralequation}{Eqs.}{Eqs.}
\crefformat{pluralequation}{Eqs.~(#2#1#3)}

\usepackage{xcite}
\usepackage{xr}

\makeatletter
\newcommand*{\addFileDependency}[1]{
  \typeout{(#1)}
  \@addtofilelist{#1}
  \IfFileExists{#1}{}{\typeout{No file #1.}}
}
\makeatother

\DeclareMathOperator\sign{sign}

\DeclareMathOperator\Lm{Lm}
\DeclareMathOperator\Prob{P}

\newcommand{\avg}[1]{\left\langle{#1}\right\rangle}
\newcommand{\E}[1]{\left\llbracket #1 \right\rrbracket}

\newcommand{\half}{\frac{1}{2}}

\newcommand{\wh}[1][x]{\widehat{#1}}

\newcommand{\lrbracket}[1]{\left(#1 \right)}
\newcommand{\lrb}[1]{\lrbracket{#1}}

\newcommand{\lrsqbracket}[1]{\left[#1\right]}
\newcommand{\lrsb}[1]{\lrsqbracket{#1}}

\newcommand{\lrcurlbracket}[1]{\left\{#1\right\}}
\newcommand{\lrcb}[1]{\lrcurlbracket{#1}}

\newcommand{\explrbr}[1]{\exp{\lrbracket{#1}}}

\newcommand{\Lmbr}[1]{\Lm\lrb{#1}}

\newcommand{\diff}[2][]{\frac{d #1}{d #2}}

\newcommand{\funcdiff}[2][]{\frac{\delta #1}{\delta #2}}

\newcommand{\funcddiff}[3][]{\frac{\delta^2 #1}{\delta #2 \delta #3}}

\newcommand{\be}{\begin{equation}}
\newcommand{\ee}{\end{equation}}
\newcommand{\bd}{\begin{displaymath}}
\newcommand{\ed}{\end{displaymath}}
\newcommand{\BE}{\begin{eqnarray}}
\newcommand{\EE}{\end{eqnarray}}

\begin{document}
    \title{Generalised Lotka--Volterra model with hierarchical interactions}
    
    \author{Lyle Poley}
  \affiliation{Theoretical Physics, Department of Physics and Astronomy, School of Natural Science, The University of Manchester, Manchester M13 9PL, UK}
    \author{Joseph W. Baron}
    \affiliation{Instituto de F{\' i}sica Interdisciplinar y Sistemas Complejos IFISC (CSIC-UIB), 07122 Palma de Mallorca, Spain}
    \author{Tobias Galla}
    \affiliation{Instituto de F{\' i}sica Interdisciplinar y Sistemas Complejos IFISC (CSIC-UIB), 07122 Palma de Mallorca, Spain}
  \affiliation{Theoretical Physics, Department of Physics and Astronomy, School of Natural Science, The University of Manchester, Manchester M13 9PL, UK}

    \date{\today}
    \begin{abstract}
        In the analysis of complex ecosystems it is common to use random interaction coefficients, often assumed to be such that all species are statistically equivalent. In this work we relax this assumption by choosing interactions according to the cascade model, which we incorporate into a generalised Lotka--Volterra dynamical system. These interactions impose a hierarchy in the community. Species benefit more, on average, from interactions with species further below them in the hierarchy than from interactions with those above. Using dynamic mean-field theory, we demonstrate that a strong hierarchical structure is stabilising, but that it reduces the number of species in the surviving community, as well as their abundances. Additionally, we show that increased heterogeneity in the variances of the interaction coefficients across positions in the hierarchy is destabilising. We also comment on the structure of the surviving community and demonstrate that the abundance and probability of survival of a species is dependent on its position in the hierarchy.
    \end{abstract}

    \maketitle

    \section{Introduction}
	The study of complex ecosystems has been an active research area in theoretical ecology since the 1970s, when evidence emerged of a sharp transition from stability to instability in model ecosystems of increasing complexity \cite{GARDNER1970,MAY1972}. By suggesting fundamental limits on the size, connectedness, and interaction variability of a stable ecosystem, these results appeared to contradict the prevailing ecological view of the time. Many ecological networks are both large and highly interconnected \cite{Dunne12917,Food_Web_Structure}, leading to an \textit{a priori} expectation that a more complex and well-connected ecosystem ought to be more stable than simpler, more sparsely connected counterpart \cite{MacArthur_1955,McCann2000}. Such an apparent contradiction has led to an increasingly detailed and nuanced search for consistent definitions of ecological stability and complexity across theory and experiment \cite{Grimm1997,Stability_Complexity_review_allesina,Landi_review,Jacquet2016,McCann2000}.
	
    One powerful theoretical tool for analysing which factors contribute to the stability of a large system of many interacting constituents, such as a complex ecosystem, is random matrix theory (RMT). This is the method that was employed by Robert May in his seminal work \cite{MAY1972}. May started from the Jacobian of a hypothetical ecosystem, linearised about an `equilibrium'. He assumed that the entries of this Jacobian matrix were i.i.d. random variables. This allowed for the deduction of a stability criterion using Girko's well-known circular law \cite{girko1985circular}. May's initial and somewhat austere model has been extended in recent years, accounting for additional features of ecosystems such as food web structure \cite{Grilli2016,Allesina2012}, spatial dispersal \cite{Baron2020Diffusion,gravel2016stability}, alternative interpretations of `interaction strength' \cite{gross2009generalized,berlow2004interaction} and varying self-regulation \cite{barabas2017self}.
    
    Despite providing a great deal of insight, RMT suffers from one crucial drawback: there is no guarantee that the random matrix under consideration corresponds to the Jacobian matrix of a feasible equilibrium \cite{stone2018feasibility,gibbs2018effect}, that is, an equilibrium at which species abundances are non-negative. To remedy this, recent works \cite{Galla_2018,baron2022non,BuninOpenLVDynamics, biroli2018marginally,altieri2021properties} have instead studied the stability of complex ecosystems using the generalised Lotka--Volterra equations, which produce feasible equilibria by construction.
    
    In this work, we extend a recent study \cite{Barabas_Cascade}, which used RMT to determine the impact of hierarchical interactions (known as the cascade model \cite{Food_Web_Structure,LVCM}) on stability. Here, we incorporate such hierarchical interactions in the generalised Lotka--Volterra equations. We obtain analytical results for both stability and the composition of surviving communities using techniques from statistical physics and the theory of disordered systems, specifically dynamic mean-field theory \cite{De_Dominics1978,MSR_formalism}. Our approach has several advantages. Foremost, as mentioned above, the equilibria we study are feasible by construction. Further, we are able to study the effects of hierarchical interactions not only on stability, but also on the emergent properties of the community, such as species abundances and survival probabilities. Finally, we are also able to identify not only \textit{when} the ecosystem becomes unstable, but also the nature of the instability.
	
	More specifically, we show that increasing the severity of hierarchical interaction in the community is a stabilising force, as is increasing the proportion of interactions which are of predator-prey type (in agreement with, for example \cite{BuninOpenLVDynamics,Tang_correlation,Stability_Complexity_review_allesina,LVCM}). We also demonstrate that larger asymmetry in the variance of species' interactions decreases stability. Further, we look at the properties of stable equilibria produced in such a hierarchical community. We find that the presence of hierarchical interaction leads to communities with non-Gaussian species abundance distributions, and that species lower in the hierarchy (i.e. species that benefit less from interactions) are both less likely to survive and less abundant. 
	
    This remainder of the paper is structured as follows: In Section \ref{section:model}, we describe the generalised Lotka--Volterra model with hierarchical (cascade) interactions. In Section \ref{section:dmft}, we outline how dynamic mean-field theory is used to trade the set of coupled differential equations with random interaction coefficients constituting the generalised Lotka--Volterra model for a smaller number of statistically equivalent coupled stochastic differential equations. We then analyse the fixed-point solution found in \cref{section:fixedpoint}, noting that the fixed-point equations were also obtained with the cavity method in Refs.~ \cite{Barbier2156} and \cite{Barbier_cavity_method} [\cref{eq:FPeqnsGeneral}], but no stability analysis was performed in these references. We then discuss the effect of hierarchical interactions on the distribution of species abundances and the fraction of surviving species in \cref{section:abundance distributions}. Finally, in \cref{section:Stability}, we study the stability of the fixed-point solution before concluding in \cref{section:discussion}.

\section{Model Definition}\label{section:model}
\begin{figure*}
    \centering
    \includegraphics[width=\textwidth]{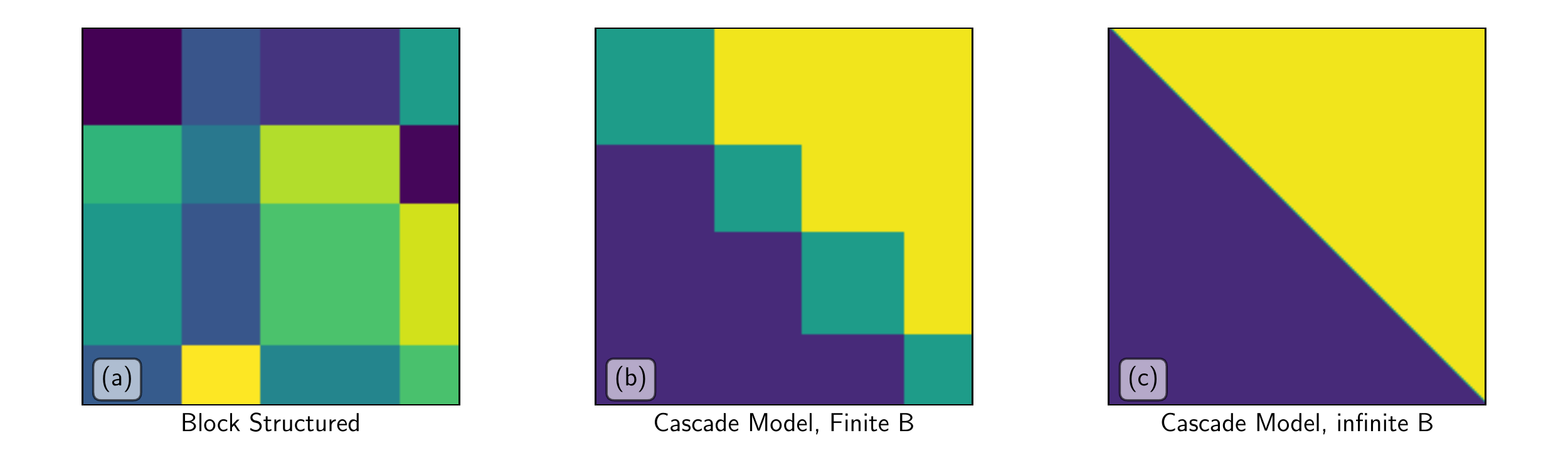}
    \caption{Illustration of the block-structured interaction matrix $A^{ab}_{ij}$. The figure shows the mean value (indicated by color) of interaction matrix elements $A^{ab}_{ij}$ in three cases. Panel (a) is a general block structured interaction matrix with $B = 4$ subcommunities. Panel (b) shows an example of an interaction matrix with cascade-model statistics, again with $B = 4$. All blocks above the diagonal blocks have the same mean and variance, and similarly for blocks below the diagonal. Panel (c) is for a model with an infinite number of sub-communities ($B \to \infty$) and cascade statistics [see \cref{eq:cascadeDefn}].}
    \label{fig:interaction_matrix}
\end{figure*}
\subsection{General Block-Structured Interactions} \label{section:blockstructureddef}

  We consider a community of $N$ species, partitioned into $B$ distinct sub-communities, which we index with $a=1,\dots,B$. Sub-community $a$ contains $N^a$ species, so that $N=\sum_a N^a$.  We first describe a more general model before introducing the specific case of the cascade model.
  
  We denote the abundance of the $i$-th species in group $a$ by $x^a_i(t)$, so that $i=1,\dots, N^a$ when we are referring to species in group $a$. Species abundances evolve according to generalised Lotka--Volterra (GLV) dynamics  \cite{Galla_2018,BuninOpenLVDynamics,Laura_2020,allesina_GLVTour,Kondoh_GLV,Barbier2156}

\begin{align}
    	\dot x_i^a(t) = x_i^a(t) \lrsb{1 - x^a_i(t) + \sum_{b = 1}^{B} \sum_{j = 1}^{N^b} A_{ij}^{ab} x_j^b(t)} \label{eq:GLV},
\end{align} 

    where the block-structured interaction matrix $(A^{ab})_{ij}$ dictates the influence of the $j$-th species in group $b$ on the $i$-th species in group $a$. The coefficients $A^{ab}_{ij}$ are correlated random variables with statistics to be specified (which may vary between between blocks). The diagonal elements $A^{aa}_{ii}$ are set to zero. This general setup is illustrated in Fig.~\ref{fig:interaction_matrix}(a).
    
    The statistics of the interactions between species depend only on their respective sub-communities. We write

    \begin{align}\label{eq:A_z}
    	A^{ab}_{ij} = \frac{\mu^{ab}}{N} + \frac{\sigma^{ab}}{\sqrt{N}}w^{ab}_{ij},
    \end{align}
    where the $w_{ij}^{ab}$ are random variables with the following first and second moments, 	
  \begin{align}
  \label{eq:statistics_z}
    	\overline{w^{ab}_{ij}} &= 0, \nonumber \\
    	\overline{(w^{ab}_{ij})^2} &= 1, \nonumber \\ \overline{w^{ab}_{ij}w^{ba}_{ji}} &= \gamma^{ab} . 
    \end{align}
We have indicated an the average over realisations of the matrix $w^{ab}_{ij}$ with an overbar. 

The model is thus fully specified by the parameters $\mu^{ab}, \sigma^{ab}$ and $\gamma^{ab}$. More specifically, $\mu^{ab}$ is the average influence of species in group $b$ on those in group $a$, $(\sigma^{ab})^2$ is the variance of those interactions and $\gamma^{ab} \in [-1, 1]$ is a correlation coefficient controlling the proportion of interactions between species in sub-communities $a$ and $b$ which are of a predator-prey type (i.e.  $A^{ab}_{ij}A^{ba}_{ji} < 0$). The factors of $1/N$ in Eq.~(\ref{eq:A_z}) ensure a sensible thermodynamic limit, $N\to\infty$ (see e.g. \cite{mezard1987}). 

\subsection{The Cascade Model}
The cascade model is obtained from Eqs.~(\ref{eq:A_z}) and (\ref{eq:statistics_z}) through a specific choice of model parameters $\mu^{ab}, \sigma^{ab}, \gamma^{ab}$. We imagine a ranking of the $B$ sub-communities in which, on average, species with higher rank gain more from lower ranked species than vice versa. Hence species higher up in the hierarchy are increasingly biased towards success.

Similar to Ref.~\cite{Barabas_Cascade}, we achieve this with the choice

    \begin{align}
        \begin{aligned}
        	\mu^{ab} = 
        	\begin{cases}
        		\mu - \nu,\hspace{-1mm} &a < b \\
        		\mu,\hspace{-1mm} &a = b \\
        		\mu + \nu,\hspace{-1mm} &a > b
        	\end{cases},\hspace{-1mm}
        \end{aligned} 
        &&
        \begin{aligned}
        	\sigma^{ab} = 
        	\begin{cases}
        		\sigma/\rho,\hspace{-1mm} &a < b \\
        		\sigma,\hspace{-1mm} &a = b \\
        		\sigma\rho,\hspace{-1mm} &a > b
        	\end{cases}, \hspace{-1mm}
        \end{aligned}
        &&
        \begin{aligned}
        	\gamma^{ab} = \gamma,
        \end{aligned}\label{eq:cascadeDefn}
    \end{align}
    where we mostly focus on the case $\nu > 0$. In this case species $a = 1$ is lowest in the hierarchy and species $a = B$ is highest. If we make the transformation $\nu \to -\nu$ and $\rho \to 1/\rho$, the ranking of species is reversed and we obtain identical results. We also require $\rho > 0$, ensuring the variance of all interactions remains positive. An example of the structure of the interaction matrix resulting from this choice is illustrated in \cref{fig:interaction_matrix} (b).

    The model parameters ($\mu, \nu, \sigma, \rho, \gamma$) can be separated into two sets: $\nu$ and $\rho$ describe the hierarchy of species, whereas $\mu, \sigma$, and $\gamma$ describe overall statistics. The parameter $\nu$ is a measure of the strength of the hierarchy. A larger value of $\nu$ increases the average benefit to species higher in the hierarchy, and correspondingly decreases the average benefit to species lower in the hierarchy. The parameter $\rho$ is a measure of the disparity between the variances of interactions $(\sigma^{ab})^2$ and $(\sigma^{ba})^2$ for $a\neq b$. The parameter $\mu$ characterises the mean interaction strength across all pairs of species in the community, $\sigma$ is a measure of the overall variability of interactions, and $\gamma$ can be related to the proportion of interactions that are of predator-prey type, $p$, via $\gamma = \cos(\pi p) + \mathcal{O}\lrb{1/\sqrt{N}}$ [see Section S2 of the Supplemental Material (SM) for details]. We note that if $\nu = 0$ and $\rho = 1$, there is no distinction between any two positions in the hierarchy and all species are statistically equivalent, this is the case considered in \cite{Galla_2018,BuninOpenLVDynamics}.
    
    It is important to note that we distinguish between the notions of hierarchy, controlled by $\nu$, and the proportion of predator-prey pairs in the model, controlled by $\gamma$. Hierarchy is present so long as $\nu \neq 0$. There is then a natural ordering to the subgroups of species. The choice $\nu>0$ implies an average benefit of species higher in the hierarchy relative to those lower in the hierarchy. The proportion of predator-prey pairs, on the other hand, is not affected by hierarchy (i.e. a non-zero value of $\nu$) to leading order in $N$ (see Section S2 in the SM). This is due to the $1/N$ scaling of the average value of $A_{ij}^{ab}$ in \cref{eq:A_z} compared to the $1/\sqrt{N}$ scaling of the random term.
\section{Dynamic mean-field theory}\label{section:dmft}
        \begin{figure*}
        	\centering
        	\includegraphics[width=0.8\textwidth]{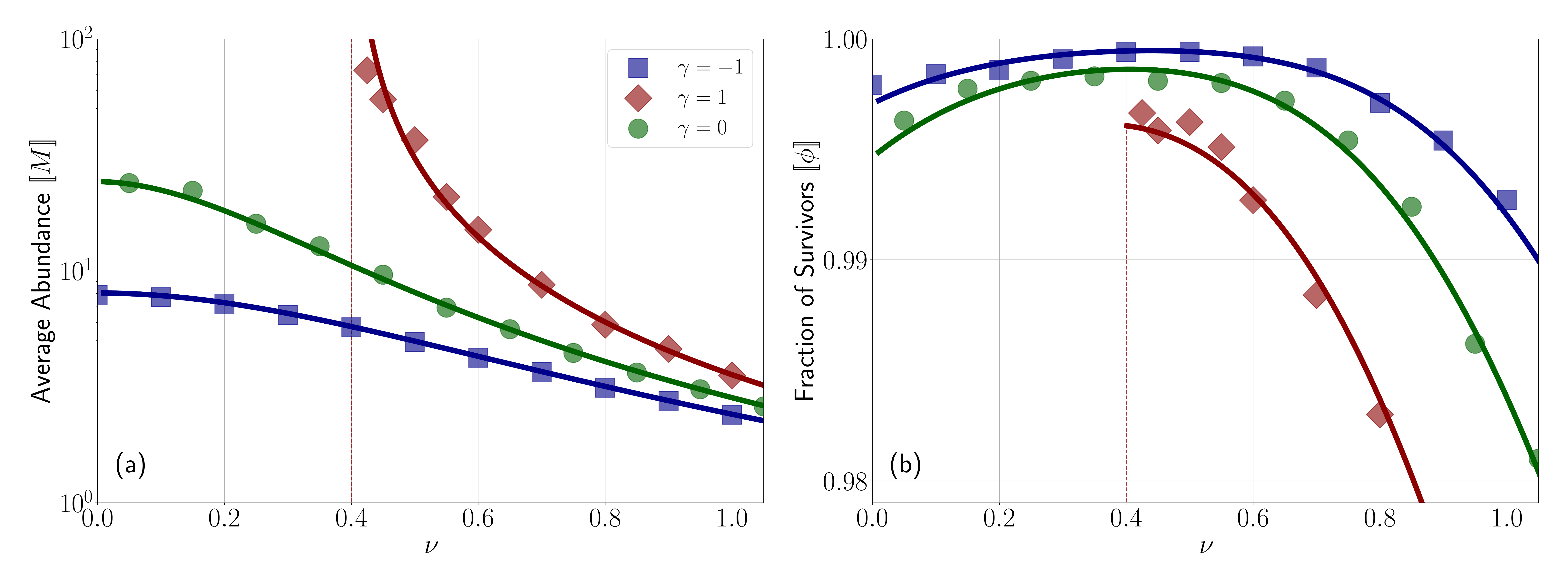}
        	\caption{Variation of average abundance $\E{M}$ and the fraction of surviving species $\E{\phi}$ with $\nu$ (the strength of the hierarchy). Panel (a): Dependence of average abundance on $\nu$. The vertical dashed line indicates a divergence in average abundance when $\gamma = 1, \nu \approx 0.4$. The system does not converge for $\nu \lesssim 0.4$. Panel (b): Dependence of the fraction of surviving species on $\nu$. When abundances diverge (for $\gamma = 1, \nu \leq 0.4$), the fraction of surviving species cannot be meaningfully calculated and so the lower curve does not extend to the left of the vertical line. Markers are data from the numerical integration of Eqs.~(\ref{eq:GLV}), with N = 250 species, averaged over 40 runs, and lines are predictions from theory [\cref{eq:phi,eq:M}, with $u, \Delta_0, \Delta_1$ determined from \cref{eq:FPeqnsCascade}]. The model parameters used in both plots are $\sigma = 0.3, \rho = 1.5, \mu = 1.0$.}
        	\label{fig:FPEqns_work}
        \end{figure*}

   We use dynamic mean-field theory \cite{BiolHandbookGF,1overf_noise,Ganguli_deep_learning_review,Galla1232,baron2021eigenvalues, De_Dominics1978,MSR_formalism,SpinGlassesBook} in order to analyse the stability and community properties of the GLV system in \cref{eq:GLV}. Ultimately our analysis will focus on models with cascade interactions, determined by the parameters $\mu, \nu, \sigma, \rho$ and $\gamma$. However, for now, we return to the more general block-structured general case discussed in Section \ref{section:blockstructureddef} [Fig.~\ref{fig:interaction_matrix}(a)].
    
  The analysis involves taking the limit $N\to\infty$ whilst keeping the ratios $n^a \equiv N^a/N$ constant, such that
  \begin{align}
      \sum_b n^a = 1.
  \end{align}
 That is to say, each sub-community contains a large (formally infinite) number of species, but the proportions of species in each group remain fixed as $N$ is varied.
   
  The calculation closely follows the lines of \cite{1overf_noise,BiolHandbookGF,Galla_2018,Laura_2020}, with modifications made to account for the block structure of the interactions in the community. Details are given in Section S3 of the SM. 
   
  The dynamic mean-field approach results in the reduction of the initial set of coupled ordinary equations with random coefficients [given in Eq.~(\ref{eq:GLV}), and with $N\to\infty$] to a set of $B$ stochastic integro-differential equations, one for each sub-community. These describe the `typical' time evolution for the abundance of species in the different groups, $x^a(t)$. Carrying out the steps in Section S2 of the SM, we arrive at the following effective dynamics for species in sub-community $a$
\begin{multline}
    	\dot x^a(t) = x^a(t)\Bigg\{1 - x^a(t) + \sum_{b = 1}^B n^b \mu^{ab}M^b(t)  \\
    	\hfill + \sum_{b = 1}^B n^b\gamma^{ab}\sigma^{ab}\sigma^{ba}\int_0^t dt' G^b(t, t')x^a(t') 
    	 + \eta^a(t)\Bigg\}. \label{eq:effectiveDynamics}
\end{multline}

The variables $\{\eta^a(t)\}$ are coloured Gaussian noise terms with the following statistics
    \begin{align}
    	\avg{\eta^a(t)} &= 0, \nonumber \\
    	\avg{\eta^a(t)\eta^b(t')} &= \delta^{ab} \sum_{c = 1}^B (\sigma^{ac})^2 n^c C^c(t, t'),
    \end{align}
where we have used $\avg{\cdots}$ to denote averages over realisations of the effective dynamics in Eq.~(\ref{eq:effectiveDynamics}), i.e. over realisations of the noise $\{\eta^a(t)\}$.

The macroscopic statistics of community $a$ are given by 
\begin{align}\label{eq:macrostat}
    M^a(t) &\equiv \avg{x^a(t)}, \nonumber \\
    C^a(t,t') &\equiv \avg{x^a(t)x^a(t')}, \nonumber \\
    G^a(t, t') &\equiv \avg{\funcdiff[x^a(t)]{\eta^a(t')}}.
\end{align}
The above quantities describe the average abundance of a species in sub-community $a$, the auto-correlations (in time) of a species abundance in the community, and the response to perturbations, respectively. The solution of Eqs.~(\ref{eq:effectiveDynamics})-(\ref{eq:macrostat}) determines the macroscopic statistics $M^a(t)$, $C^a(t,t')$ and $G^a(t, t')$ self-consistently.

\section{Fixed Point Equations}\label{section:fixedpoint}

\subsection{General Block-Structured Interactions}

    We now assume that the system reaches a fixed point such that $x^a(t) \to x^a_*$ as $t \to \infty$, where $x^a_*$ is a static random variable. We write
    \begin{align}
        M^a &\equiv \avg{x^a_*},
        \nonumber\\ 
        q^a &\equiv \avg{(x^a_*)^2}, \label{eq:fixpointdef1}
    \end{align}
    for the first and second moments of the asymptotic abundances in each group of species $a$. The noise term $\eta^a(t) \to \eta^a_*$ also asymptotically loses its time dependence and we write
    \begin{align}
    \eta^a_* = z^a\sqrt{\sum_b (\sigma^{ab})^2 n^b q^b},\label{eq:fixpointdef2}
    \end{align}
    where the $z^a$ are independent zero-mean static Gaussian random variables with unit variance. In this fixed-point regime the response function $G^a(t,t') \to G^a(t - t')$ only depends on time differences $\tau=t-t'$, and causality dictates that $G^a(\tau) = 0$ for $\tau < 0$. We then define the integrated response function for sub-community $a$
    \begin{align}
        \chi^a \equiv \int_0^\infty d\tau\ G^a(\tau). \label{eq:fixpointdef3}
    \end{align}
    With \cref{eq:fixpointdef1,eq:fixpointdef2,eq:fixpointdef3} in mind, we find that the nontrivial fixed point of Eq.~(\ref{eq:effectiveDynamics}) is described by 
\begin{align}
    	x^a_* = \max\lrb{0,  
    		\frac{1 + \sum_b \mu^{ab} n^b M^b + z^a\sqrt{\sum_b (\sigma^{ab})^2 n^b q^b}}{1 - \sum_b \gamma^{ab} \sigma^{ab} \sigma^{ba} n^b \chi^b} }. \label{eq:staticx}
\end{align}%
A similar expression was found for models without hierarchical structure in \cite{1overf_noise,Galla_2018,bunin2016interaction,Laura_2020}. We note that depending on the value that the random variable $z^a$ takes, some species will be extinct at the fixed point ($x^a_*=0$), whereas others will have positive abundance.

The average over realisations of the effective process, $\avg{\dots}$, is now an average over the static Gaussian random variables $z^a$ at the fixed point. This allows us to obtain self-consistent conditions for the statistics of $x_*^a$. We find (see Section~S4.1 of the SM)
\begin{align}
    \chi^a u^a &= w_0(\Delta^a), \nonumber \\
    M^a u^a &= w_1(\Delta^a)\sqrt{\sum_b (\sigma^{ab})^2 n^b q^b}, \nonumber \\
    q^a \lrb{u^a}^2 &= w_2(\Delta^a) \sum_b (\sigma^{ab})^2 n^b q^b, \label[pluralequation]{eq:FPeqnsGeneral}
\end{align}
where we have abbreviated
\begin{align}
    u^a &= 1 - \sum_b \gamma^{ab} \sigma^{ab} \sigma^{ba} n^b \chi^b, \nonumber \\
    \Delta^a &=\frac{1 + \sum_b \mu ^{ab} n^b M^b}{\sqrt{\sum_b (\sigma^{ab})^2 n^b q^b}}, \label[pluralequation]{eq:FPeqnsGeneral2}
\end{align}
and where we have defined the functions
\be
w_k(\Delta) = \frac{1}{\sqrt{2\pi}}\int_{-\infty}^\Delta dz\ e^{-\half z^2} (z - \Delta)^k,
\ee
for $k\in\{0,1,2\}$. \cref{eq:FPeqnsGeneral} are equivalent to those in section IV.1 of the Supplementary Information of Ref.~\cite{Barbier2156}, derived with the cavity method. They also reduce to the fixed point equations for the model without hierarchy found in Refs. \cite{BuninOpenLVDynamics,Galla_2018} if one takes $\mu^{ab} = \mu/B, \sigma^{ab} = \sigma/\sqrt{B}$ and $\gamma^{ab} = \gamma$.

    \cref{eq:FPeqnsGeneral} can be solved numerically for the quantities $\chi^a, M^a, q^a$ as functions of the model parameters $\mu^{ab}, \sigma^{ab}, \gamma^{ab}$ and $n^a$. We note that the fraction of surviving species in sub-community $a$, $\phi^a$, is given by 
    \be
    \phi^a=w_0(\Delta^a).
    \ee
    Hence the solution of \cref{eq:FPeqnsGeneral} also provides the fraction of surviving species in each  sub-community.
    
    For further analysis, it is useful to introduce the following average over communities
    \be
    \E{X}_B = \sum_{a=1}^B n^a X^a
    \ee
    for a quantity $X^a$ defined in each community. The overall fraction of surviving species in the system is then  $\E{\phi}_B$, and the average abundance per species is $\E{M}_B$. 

    \begin{figure*}
    	\centering
        \makebox[\textwidth][c]{\includegraphics[width=\textwidth]{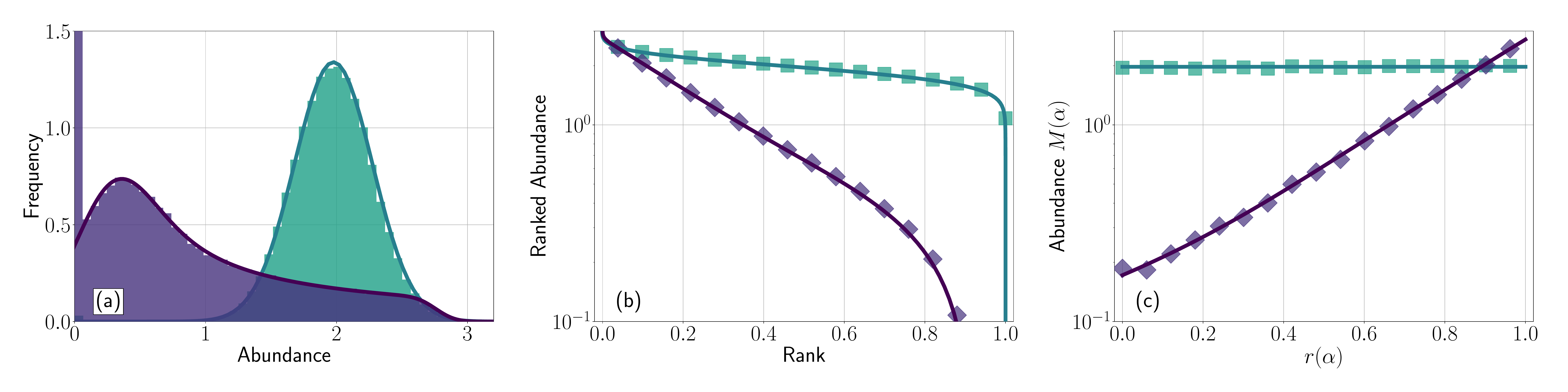}}%
    	\caption{Abundance distributions for the cascade model with and without hierarchical interaction. All plots are for $\mu = 0.5,\ \sigma = 0.15,\ \gamma = -0.3$. Panel (a): Two SADs are plotted, the taller one is for the model without hierarchy ($\nu = 0, \rho = 1$), and is a clipped Gaussian. The flatter distribution in panel (a) is for $\nu = 1.5, \rho = 2/3$, bars are from simulation. Panels (b) and (c): Corresponding RADs and HADs. Square markers are from simulations for $\nu = 0, \rho = 1$, and diamond markers from simulations for $\nu = 1.5, \rho = 2/3$. Simulations are for $N = 1000$ species, averaged over $200$ runs. Lines are from the theory.}
    	\label{fig:abundance_distributions}
    \end{figure*}

\subsection{Cascade Model} \label{section:for the cascade model}
    
    We now analyse the fixed-point solution for a community with cascade-model interactions, for which the parameters $\mu^{ab}, \sigma^{ab}, \gamma^{ab}$ are as in Eq.~(\ref{eq:cascadeDefn}). It is convenient to introduce $\alpha=a/B$, and to work in the limit $B\to\infty$, the resulting interaction matrix is illustrated in Fig.~\ref{fig:interaction_matrix}(c). The index $\alpha \in [0, 1)$ is now continuous, and the constraint $\sum_{b=1}^B n^b=1$ becomes $\int_0^1 d\alpha\, n(\alpha)=1$ (see also Section S4.1 of the SM). Quantities such as $M^a, \phi^a, \Delta^a$ etc. are now functions of $\alpha$, and averages over the index $a$ are integrals, which we write as follows
    \begin{align}
        \E{X}\equiv \lim_{B \to \infty}\E{X}_B  = \int_0^1 d\alpha~ n(\alpha) X(\alpha),
    \end{align}
    dropping the subscript $B$ in the limit. For example, the first relation in \cref{eq:FPeqnsGeneral2} is now
    \begin{align}
        u(\alpha) = 1 - \gamma\sigma^2\E{\chi},
    \end{align}
    informing us that $u$ has no dependence on $\alpha$ in the cascade model when $B \to \infty$.
    
   Recalling the definition of $\Delta^a$ in the second relation in Eq.~(\ref{eq:FPeqnsGeneral2}) we also introduce the following quantities  
    \begin{align}
        \Delta_0 \equiv \Delta(\alpha=0),\hspace{1cm} \Delta_1 \equiv \Delta(\alpha=1).
    \end{align}
   To proceed, we first find expressions for the quantities $u, \Delta_0, \Delta_1$, in terms of $\nu, \sigma, \rho, \gamma$ using \cref{eq:FPeqnsGeneral}. Then, we will find expressions for $\E{M}$ and $\E{\phi}$ [we also discuss expressions for the full functions $M(\alpha), \phi(\alpha)$ in \cref{section:abundance distributions}] in terms of $u, \Delta_0, \Delta_1$. In turn, this will enable us to calculate the macroscopic statistics of the system for any $\mu, \nu, \sigma, \rho, \gamma$. Details of all remaining calculations in this section can be found in Section S4 of the SM.
    
    Manipulation of \cref{eq:FPeqnsGeneral} reveals that $u$ satisfies the following [see Eqs.~(S49) and (S57) in Section S4 of the SM]
    \begin{align}
        u^2 &= \ell\sigma^2\int_0^1d\alpha~n(\alpha)w_2[\Delta(\alpha)], \nonumber \\
        u(1 - u) &= \gamma\sigma^2\int_0^1d\alpha~n(\alpha)w_0[\Delta(\alpha)], \label[pluralequation]{eq:urelations}
    \end{align}
    where $\ell$ is the logarithmic mean \cite{Log_mean} of $\rho^2$ and $1/\rho^2$
    \begin{align}
        \ell \equiv \frac{\rho^2-1/\rho^2}{\ln \rho^2 - \ln 1/\rho^2}, \label{eq:ell}
    \end{align}
    which satisfies $\ell \geq 1$, with equality only if $\rho = 1$. We now show that the integrals in \cref{eq:urelations} can be written explicitly in terms of only $u, \Delta_0, \Delta_1$. One achieves this by finding the following separable differential equation for $\Delta(\alpha)$ [Eq.~(S67) in Section S4 of the SM]
    \begin{align}
        \diff[\Delta]{\alpha} = \frac{n(\alpha)}{\theta(\Delta,u)}, \label{eq:DeltaODE}
    \end{align}
    with initial condition $\Delta(0) = \Delta_0$ and with
    \begin{align}
        \theta(\Delta,u) \equiv \frac{u^2}{2 u \nu w_1(\Delta) - \ell \sigma^2 \ln\rho^2 \Delta w_2(\Delta)}. \label{eq:thetadefn}
    \end{align}
    Self-consistently, the solution to \cref{eq:DeltaODE} will further have to satisfy the constraint $\Delta(1) = \Delta_1$.

    One then uses \cref{eq:DeltaODE} to change variables in the integrals in \cref{eq:urelations}, finding
    \begin{align}
        \int_0^1d\alpha~n(\alpha)w_2[\Delta(\alpha)] &= \int_{\Delta_0}^{\Delta_1}d\Delta~\theta(\Delta, u)w_2(\Delta), \label{eq:changeOfVariables}
    \end{align}
    and similarly for the integral over $w_0[\Delta(\alpha)]$. If we also perform the same change of variables on the condition $\int_0^1d\alpha~n(\alpha) = 1$ then we arrive at the following simultaneous equations for $u, \Delta_0$ and $\Delta_1$
    \begin{align}
        u^2 &= \ell \sigma^2 \int_{\Delta_0}^{\Delta_1}d\Delta~\theta(\Delta,u)w_2(\Delta),  \nonumber \\
        u(1-u) &= \gamma \sigma^2 \int_{\Delta_0}^{\Delta_1}d\Delta~\theta(\Delta,u)w_0(\Delta), \nonumber \\ 
        1 &= \int_{\Delta_0}^{\Delta_1}d\Delta~\theta(\Delta,u). \label[pluralequation]{eq:FPeqnsCascade}
    \end{align}
    Given the parameters $\nu, \sigma, \rho, \gamma$, \cref{eq:FPeqnsCascade} can be solved numerically to obtain $\Delta_0, \Delta_1$ and $u$. As the parameter $\mu$ and the function $n(\alpha)$ do not appear in \cref{eq:FPeqnsCascade}, we conclude that $\Delta_0, \Delta_1$ and $u$ are independent of them. That is, $\Delta_0, \Delta_1$ and $u$ are functions of $\nu, \sigma, \rho$ and $\gamma$ only. 
    
    Once $\Delta_0, \Delta_1, u$ are determined, we calculate the average abundance and fraction of surviving species in the community with 
    \begin{subequations}
    \begin{align}
        \frac{1}{\E{M}} &= \frac{\Delta_1\rho^2 + \Delta_0}{\Delta_1\rho^2 - \Delta_0}\nu - \mu, \label{eq:M} \\
        \E{\phi} &= \frac{u(1-u)}{\gamma\sigma^2}. \label{eq:phi}
    \end{align}
    \end{subequations}
    See also Eqs.~(S77) and (S78) in Section S4 of the SM. \cref{eq:M} ceases to apply when $\nu = 0$, and a limit must be taken. Similar care has to be taken when finding $\E{\phi}$ in the limit $\gamma \to 0$. Both limits are found in Section S4.5.3 of the SM.

    Inspection of \cref{eq:M,eq:phi} reveals that, surprisingly, neither $\E{\phi}$ nor $\E{M}$ depend on $n(\alpha)$, the relative sizes of the different sub-communities $\alpha$. In fact, we will see in \cref{section:Stability,section:abundance distributions} that $n(\alpha)$ does not affect any of the properties of the community that we are interested in (see also Section S8 in the SM). One also sees that $\E{\phi}$ is independent of $\mu$.

    Our theoretical predictions for the community average species abundance $\E{M}$ and the community average survival fraction $\E{\phi}$ are verified using the results of computer simulation of Eqs.~(\ref{eq:GLV}) in Fig. \ref{fig:FPEqns_work}.

\section{Fixed Point Distributions}\label{section:abundance distributions}
\subsection{Abundance Distributions} \label{section:abundance distributions_1}
    At a stable equilibrium, we can calculate species abundance distributions (SADs) and rank abundance distributions (RADs). An SAD is obtained in simulations by binning species according to their abundances and producing a histogram of the number of species in each bin [panel (a) of \cref{fig:abundance_distributions}]. An RAD is a plot of abundance against a ranking of species from $0$ to $1$, with the highest abundance species having a rank of $1$ [panel (b) of \cref{fig:abundance_distributions}]. For reviews of both SADs and RADs see Refs. \cite{SAD_review_Matthews,SAD_review_McGill}, or see Ref. \cite{Galla_2018} for similar calculations without hierarchical interactions, and \cite{RADs_Yoshino} for RADs derived from random replicator equations rather than Lotka--Volterra. 
    
    We also introduce hierarchical abundance distributions (HADs). Similar to an RAD we rank species from $0$ to $1$, but this time such that a species with rank $r$ is higher in the hierarchy of the cascade model than $rN$ species, and lower than $(1-r)N$ species. An HAD is then a plot of abundance as a function of this hierarchical rank.
    
    To calculate SADs and RADs we use $\Prob(x|\alpha)$, the probability that a species has abundance $x$, given that it sits at position $\alpha$ in the hierarchy. This is the probability density for $x_*^a$ in \cref{eq:staticx}. One arrives at (see also Section S5 in the SM) the following clipped Gaussian distribution
\begin{align}
        \Prob(x|\alpha) = \left[1-\phi(\alpha)\right]\delta(x) + H(x)\varphi(x|\alpha),\label{eq:p_m_alpha}
\end{align}
    where $H(\cdot)$ is the Heaviside step function [$H(x)=1$ for $x\geq 0$, and $H(x)=0$ otherwise], and where the delta function $\delta(x)$ represents species that have gone extinct and hence have an abundance of $x = 0$. The function $\varphi(x|\alpha)$ is a (normalised) Gaussian in $x$ such that $\int_0^\infty dx~ \varphi(x|\alpha) = \phi(\alpha)$,
    \begin{align}
        \varphi(x|\alpha)=\frac{w_1[\Delta(\alpha)]}{M(\alpha)\sqrt{2\pi}} \exp\left\{-\frac{1}{2}\left(\frac{w_1[\Delta(\alpha)]}{M(\alpha)}x - \Delta(\alpha)\right)^2\right\}.
    \end{align}
    The SADs in \cref{fig:abundance_distributions} are given by 
    \begin{align}
        \E{\Prob(x)} \equiv \int_0^1 d\alpha~n(\alpha)\Prob(x|\alpha). \label{eq:SAD_pre_cov}
    \end{align}
    To calculate $\E{\Prob(x)}$ explicitly, first \cref{eq:FPeqnsCascade} are solved to obtain $\Delta_0, \Delta_1$ and $u$. We then re-express \cref{eq:p_m_alpha} as a function of $\Delta(\alpha)$ and use the same change of variables as in \cref{eq:changeOfVariables} to calculate the integral \cref{eq:SAD_pre_cov}, obtaining
    \begin{align}
        \E{\Prob(x)} = \lrb{1 - \E{\phi}}\delta(x) + H(x)\int_{\Delta_0}^{\Delta_1}d\Delta~\theta(\Delta, u)\varphi(x|\Delta), \label{eq:SAD}
    \end{align}
    details are found in Section S5 of the SM. Similarly to $\E{M}$ and $\E{\phi}$, the distribution $\E{\Prob(x)}$ is independent of $n(\alpha)$ (see Section S8 of the SM).
    
    In the case where $(\nu = 0, \rho = 1)$ (i.e. no hierarchy), the underlying unclipped distribution $\E{\varphi(x)} = \int_0^1d\alpha~n(\alpha)\varphi(x|\alpha)$ is itself a Gaussian distribution. However, as demonstrated in \cref{fig:abundance_distributions}(a), this simple form is lost in a hierarchical community, $\E{\varphi(x)}$ is no longer Gaussian, and is not even symmetric around its maximum. Broadly speaking, raising $\nu$ or $\rho$ lowers the modal abundance and increases spread in abundances respectively. 
    
    RADs are also calculated from the distribution $\E{\Prob(x)}$. We observe that if species are ranked on a scale of $0$ to $1$ by descending abundance, then the rank of a species with abundance $x$ is $\int_{x}^{\infty}\E{\Prob(x')}dx'$. The plot in \cref{fig:abundance_distributions}(b) shows abundance on the vertical axis, and rank on the horizontal axis. RADs are often the preferred representation of species abundances as they do not suffer from loss of information due to species binning, as SADs do \cite{magurran_2011}.  
    
    To compute a hierarchical abundance distribution we rank species on a scale $r \in [0, 1]$, a species with index $\alpha$ has rank 
    \begin{align}
        r(\alpha) = \int_0^\alpha d\alpha~n(\alpha), \label{eq:rdef}
    \end{align}
    ensuring that $rN$ species are lower in the hierarchy and $(1 - r)N$ species are higher in the hierarchy. One then produces a HAD, such as in panel (c) of \cref{fig:abundance_distributions}, parametrically, with the horizontal axis equal to the rank and the vertical axis equal to the abundance $M(\alpha)$, which is derived from \cref{eq:FPeqnsGeneral} in Section S5 of the SM and given by 
    \begin{multline}
        M(\alpha) = A\E{M}w_1[\Delta(\alpha)] \\
        \hfill \explrbr{\frac{\ell\sigma^2 \ln\rho}{u^2} \int_{\Delta_0}^{\Delta(\alpha)} d\Delta'~\theta(\Delta', u)w_2(\Delta')}, \label{eq:HAD}
    \end{multline}
    where the constant $A$ ensures that $\int_0^1 d\alpha~n(\alpha)M(\alpha) = \E{M}$, for $\E{M}$ given by \cref{eq:M}. Surprisingly, just like $\E{M}, \E{\phi}$ and $\E{\Prob(x)}$, HADs are also independent of $n(\alpha)$, this is shown in Section S8 of the SM.
    \begin{figure}
        \centering
        \includegraphics[width=0.4\textwidth]{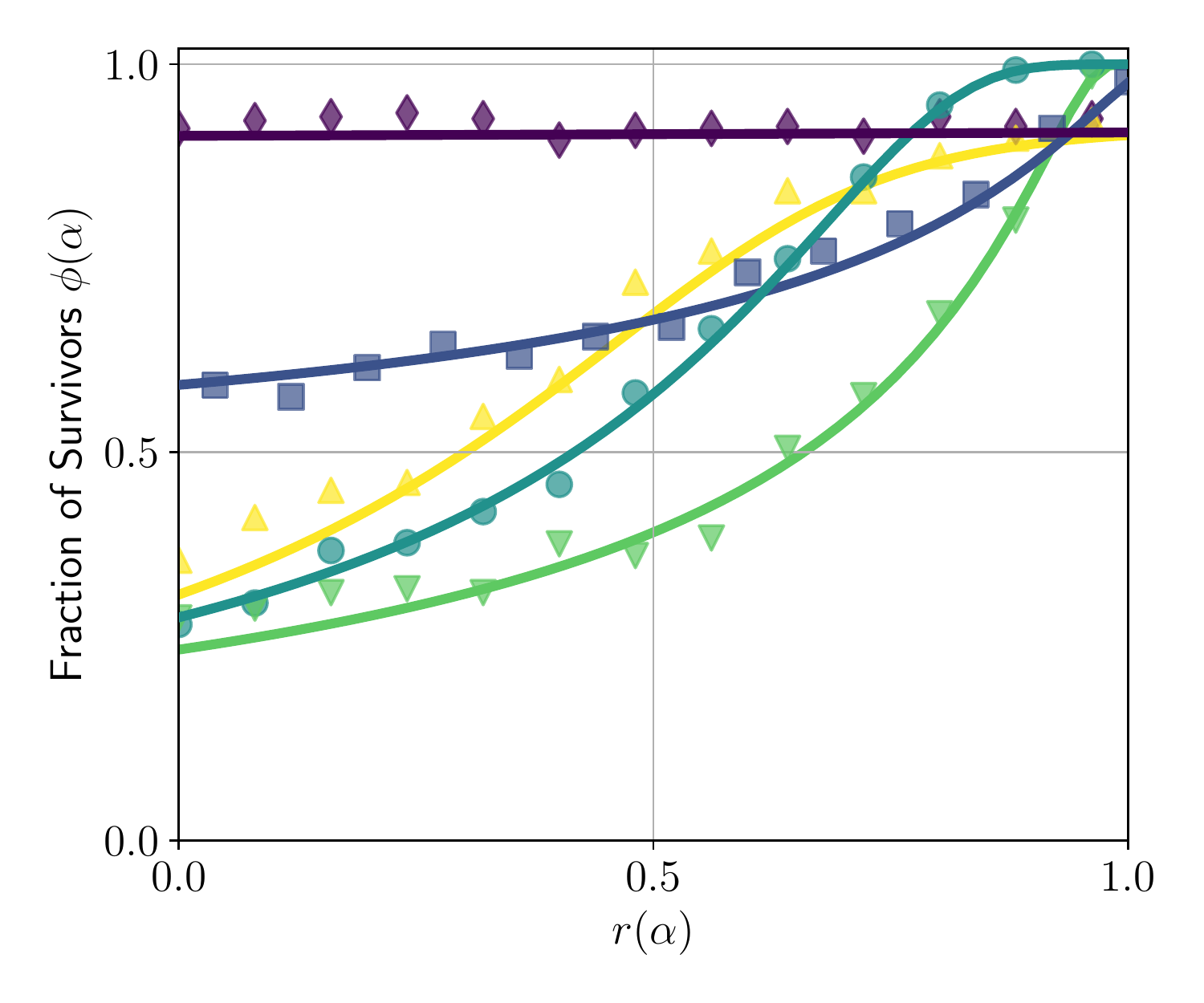}
        \caption{Survival Distributions with and without hierarchical interactions. Parameters used are $\sigma = 0.75, \gamma = -0.6$ and, from top to bottom along the left edge of the plot, $(\nu, \rho) = (0, 1), (0, 3), (3, 3), (3, 1), (3, 1/3)$. Markers are the average fraction of survivors for every fortieth species in simulations with $N = 500$, averaged over $400$ runs.}
        \label{fig:survival_distributions}
    \end{figure}
    \begin{figure*}
        \centering
        \includegraphics[width=0.8\textwidth]{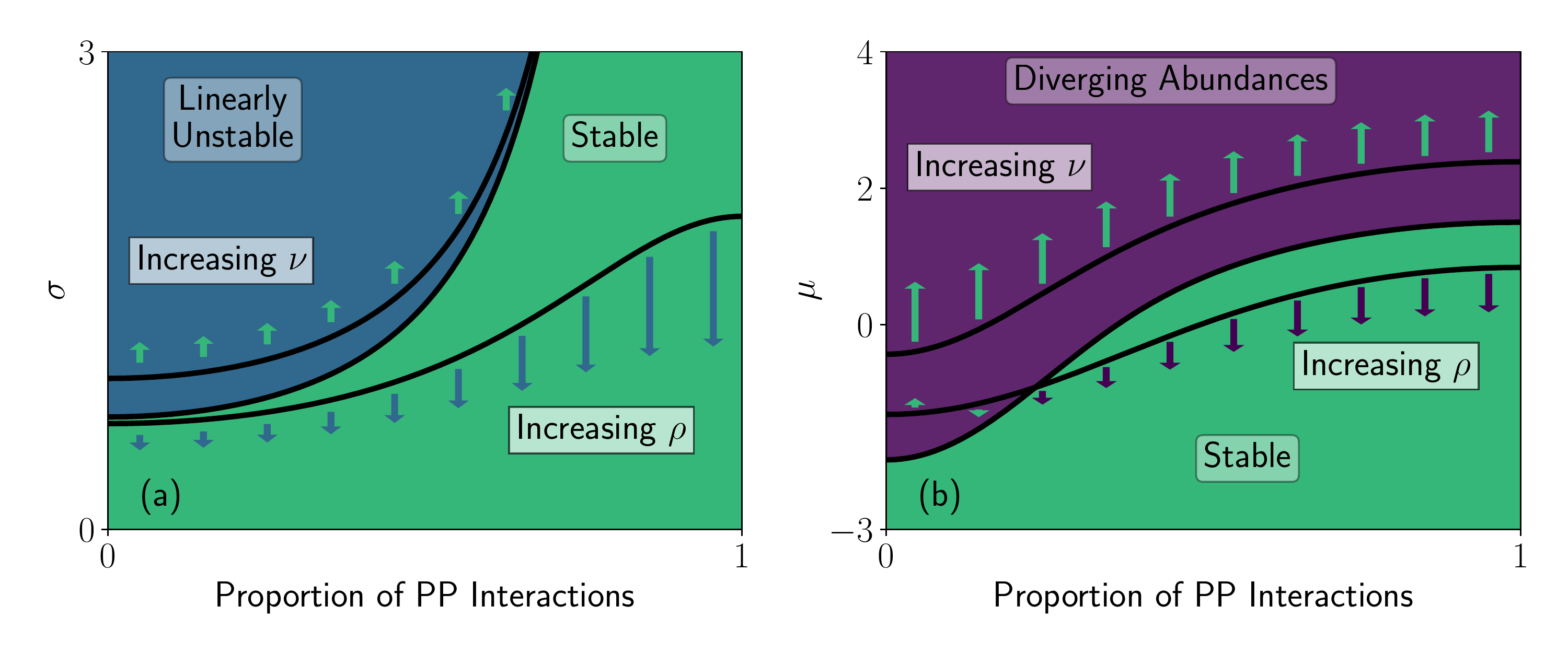}
        \caption{Phase diagrams for the cascade model (horizontal axis shows the proportion of predator-prey pairs). The onset of instability [linear instability in panel (a), and diverging abundances in panel (b)] are shown as black lines for $(\nu, \rho) = (0, 1)$, as well as for $(\nu,\rho)=(3,1)$ and $(\nu,\rho)=(0,3)$ as indicated. Stability is coloured as labelled for the case $(\nu, \rho) = (0, 1)$.  Length of arrows indicate the effect of increasing the values of $\nu$ or $\rho$ by one [$\nu\to\nu+1$, or $\rho\to\rho+1$ respectively].}
        \label{fig:gamma_stability}
    \end{figure*} 
\subsection{Fraction of Surviving Species}
    A given species' survival probability as a function of the ranking $r(\alpha)$ can be produced in a similar way to the hierarchical abundance distributions discussed in \cref{section:abundance distributions_1}. We simply replace $M(\alpha)$ by $\phi(\alpha) = w_0[\Delta(\alpha)]$. By the same reasoning, we deduce that a plot of survival probability against rank $r(\alpha)$ is independent of both $\mu$ and $n(\alpha)$ (see Section S8 of the SM). The distribution of survival probabilities is flat if $\nu = 0, \rho = 1$ and is an increasing function of $r$ in the presence of hierarchy (see Fig. \ref{fig:survival_distributions}). This is an indication that the composition of a hierarchical community before the dynamics are run is vastly different to the resulting stable community, species lower in the hierarchy are both less abundant and less likely to survive. 
    
    On further noting that the area under an HAD is $\E{M}$ and the area under the survival curve is $\E{\phi}$, we see that introducing hierarchy produces smaller communities dominated by high ranked species. Further, from \cref{fig:abundance_distributions,fig:survival_distributions}, the mean abundance and the fraction of surviving species in communities with hierarchy are mostly lower than those in models without hierarchy (flat curve in both), demonstrating that few species benefit at all from hierarchical interactions. 
\section{Stability} \label{section:Stability}

\subsection{Stability Conditions}\label{section:StabilityConditions}
    In Section \ref{section:fixedpoint}, we found the statistics of the surviving species abundances by presuming a static solution to Eq.~(\ref{eq:effectiveDynamics}). In this section, we discuss when this fixed-point solution is valid and thus under what conditions we have a stable and feasible equilibrium. As in Refs. \cite{Galla_2018,Barbier2156,BuninOpenLVDynamics,Laura_2020}, we find that instability can occur either through linear instability against small perturbations to the abundances, or through a divergence in species abundances.

    \subsubsection{Linear Instability}
    Following along the lines of \cite{1overf_noise,Galla_2018} (see Section S6 of the SM), we use a linear stability analysis to find that the system is unstable to perturbations in species abundances when 
    \begin{align}
        \sigma^2 \geq \frac{\ell}{(\ell + \gamma)^2} \frac{1}{\E{\phi}}, \label{eq:opper_transition}
    \end{align}
    where the average survival probability $\E{\phi}$ is to be determined from \cref{eq:FPeqnsCascade} and \cref{eq:phi}.
    
    Interestingly, the same criterion as in \cref{eq:opper_transition} can be obtained with the machinery of random matrix theory, noting that $N^* = \E{\phi}N$ species survive the dynamics asymptotically and reach a fixed point. The bulk of the eigenvalue spectrum of an $N^*\times N^*$ random matrix with cascade statistics as in \cref{eq:cascadeDefn} (c.f. \cref{fig:interaction_matrix}) crosses the imaginary axis precisely when \cref{eq:opper_transition} is satisfied \cite{Barabas_Cascade}. Similar observations were made in the case without hierarchy ($\nu =0$ and $\rho = 1$) in Ref. \cite{baron2022non}. A crucial difference between the dynamical and random matrix approaches is that the fraction of survivors $\E{\phi}=N^*/N$ is determined from \cref{eq:phi} in the dynamical theory, but $\E{\phi}$ is an independent parameter of the model in the random matrix approach. In particular, Eq.~(\ref{eq:opper_transition}) would lead one to conclude that $\nu$ has no effect on linear stability if a random matrix approach were used. However, as $\E{\phi}$ is itself a function of the parameters $\nu, \sigma, \rho, \gamma$, no such conclusion is drawn here.
    \begin{figure*}
        \centering
        \includegraphics[width=\textwidth]{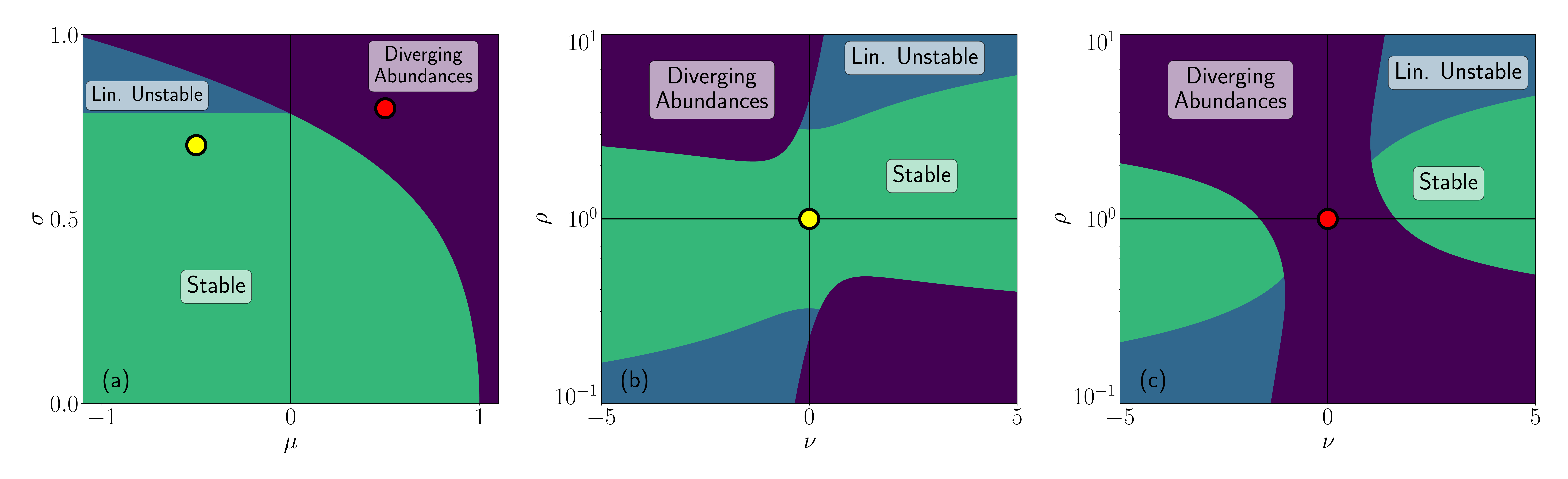}
        \caption{Hierarchical interactions and stability. Panel (a): Similar to the phase diagrams in Ref. \cite{BuninOpenLVDynamics}, we show the three possible dynamical behaviours of the system without hierarchy in the $\mu - \sigma$ plane. We have $\gamma = 0.8$ ($20\%$ predator-prey interactions), $\nu = 0$ and $\rho = 1$. In panels (b) and (c) we choose the points $\mu = -0.5, \sigma = 0.7$ [panel (b)], and $\mu = 0.5, \sigma = 0.8$ [panel (c)], and show stability in the $\nu-\rho$ plane. The left circle (yellow) in panel (a) and the circle in panel (b) are the same point in parameter space ($\mu = -0.5, \nu = 0, \sigma = 0.7, \rho = 1, \gamma = 0.8$). Similarly for the right (red) circle in (a) and the circle at the centre of panel (c) with ($\mu = 0.5, \nu = 0, \sigma = 0.8, \rho = 1, \gamma = 0.8$).}
        \label{fig:nu-rho-stability}
    \end{figure*} 
    \subsubsection{Diverging Abundances}
    To find the point at which species abundances diverge, we solve \cref{eq:FPeqnsCascade}, together with \cref{eq:M}, for the point at which $1/\E{M} = 0$. We find that abundances diverge if
    \begin{align}
        \mu &\geq \frac{\Delta_1\rho^2 + \Delta_0}{\Delta_1\rho^2 - \Delta_0}\nu . \label{eq:Mdiv}
    \end{align}
    On eliminating $\Delta_0$ and $\Delta_1$ using Eqs.~(\ref{eq:FPeqnsCascade}), \cref{eq:Mdiv} can be solved to yield the critical value of one of the parameters $(\mu, \sigma, \nu, \rho, \gamma)$ given the others. The resulting predictions for the point at which the mean abundance diverges is exact when the system is stable with respect to small perturbations, but is only approximate when the system becomes linearly unstable before the point of divergence is reached (see Section S7 of the SM for numerical justification). We include both cases in \cref{fig:nu-rho-stability}. The stability conditions \cref{eq:opper_transition,eq:Mdiv} provide a comprehensive analytical picture of stability in the cascade system.

\subsection{Phase Diagrams}
    The phase diagrams in \cref{fig:nu-rho-stability,fig:gamma_stability} illustrate the effects of hierarchical interaction ($\nu, \rho$) on stability.     
    
    When $\nu$ is sufficiently large, increasing $\nu$ typically decreases average abundances and the fraction of surviving species (see \cref{fig:FPEqns_work}), thereby pushing the system away from diverging abundances and from linear instability. Larger deviations of $\rho$ from unity push the system both towards linear instability and generally towards infinite abundances. These effects are demonstrated in panels (b) and (c) of \cref{fig:nu-rho-stability}, which indicate that a large enough value of $\rho$ will always result in an unstable system and a large enough value of $\nu$ will always result in a stable one. We also demonstrate the (separate) effects of changing $\nu$ and $\rho$ in \cref{fig:gamma_stability}. If our parameters are restricted such that either $\nu = 0$ or $\rho = 1$ [shown in panel (a) of \cref{fig:gamma_stability}], then we can analytically demonstrate the effects of varying only one parameter (either $\nu$ or $\rho$) on linear instability. Details can be found in Section S6.3 of the SM.

    Panels (b) and (c) in \cref{fig:nu-rho-stability} also reveal how the precise combination of $\nu$ and $\rho$ can affect stability and community composition. Specifically, communities with $\nu > 0$ and $\rho > 1$ tend to have lower abundances than those with $\nu > 0$ and $\rho < 1$ and are further from the diverging abundances transition. 
    
    From \cref{fig:gamma_stability}, we find that the influence of $\nu$ on linear stability is relatively small compared with its influence on the point at which abundances diverge. This is reminiscent of the fact that the overall average interaction strength $\mu$ has no effect on linear stability but affects the average abundance. \cref{fig:gamma_stability} also reveals that $\rho$ has a more dramatic effect on stability in communities with a large proportion of predator-prey pairs. In particular, any value of $\rho \neq 1$ removes the special behaviour of the model without hierarchy and all interactions of the predator-prey type ($\rho=1, \nu=0, \gamma = -1$). In this special case, there is no linear instability \cite{Galla_2018, BuninOpenLVDynamics, bunin2016interaction}, as indicated in panel (a) of Fig.~\ref{fig:gamma_stability}. Any value $\rho\neq 1$ will however lead to the possibility of instability in the system with only-predator prey interactions. The stabilising influence of $\nu$, on the contrary, is only mildly affected by the proportion of predator-prey pairs.
    
    The influence of the remaining parameters $\mu, \sigma$ and $\gamma$ on stability is relatively straightforward, and similar to the model without block structure ($\nu = 0$ and $\rho = 1$) previously studied in  \cite{Galla_2018,Laura_2020,bunin2016interaction,BuninOpenLVDynamics}, and in studies of complex ecosystems based on the spectra of random matrices \cite{MAY1972,Tang_correlation}. The parameter $\mu$ has no effect on linear stability, as \cref{eq:opper_transition} has no $\mu$ dependence. Instead, from \cref{eq:M} we see that $\mu$ increases average abundances (noting that $\Delta_0$ and $\Delta_1$ are independent of $\mu$), thereby moving the system towards instability via diverging abundances. This can be seen in panel (a) of \cref{fig:nu-rho-stability} and panel (b) of \cref{fig:gamma_stability}: a large enough value of $\mu$ will always lead to diverging abundances in the community. Increasing $\sigma$, on the other hand, pushes the system towards both linear instability and diverging abundances, as demonstrated in \cref{fig:nu-rho-stability}(a) and in \cref{fig:gamma_stability}(a). Increasing the correlation parameter $\gamma$ (i.e., decreasing the proportion of predator-prey interaction pairs), also pushes the system towards linear instability and diverging abundances, as demonstrated in \cref{fig:gamma_stability}.

\section{Discussion} \label{section:discussion}
    Our analysis of the generalised Lotka--Volterra model with cascade interactions has focused on the effect of hierarchical interactions on both stability and structure in complex ecological communities. We have extended previous work on the stabilising impact of predator-prey like relationships  \cite{Tang_correlation,Galla_2018,BuninOpenLVDynamics,Allesina2012} by considering both the average severity of the hierarchy ($\nu$) and the proportion of interactions of predator-prey type ($\gamma$) in a single dynamical model. We find that increases to both factors are stabilising. We also find that increased heterogeneity in interaction variances ($\rho$) is a destabilising force.
    
    The dynamic mean-field theory approach, unlike an approach based on the spectra of random matrices, guarantees a feasible equilibrium, and gives access to properties of the ecosystem other than stability. We find that communities with a strong hierarchy are dominated by species at the top, which are both more abundant and more likely to survive asymptotically. Further, hierarchy leads to more complex, non-Gaussian abundance distributions. 
    
    In order to find fixed point equations for the cascade model with an infinite number of trophic levels, we first considered a related and more general community, divided into a finite number of sub-communities. Fixed-point equations were then obtained, resulting in an effective abundance for a representative species $x^a(t)$ in each community. This is in contrast to most similar studies employing dynamic mean-field theory \cite{bunin2016interaction,Galla_2018,1overf_noise}, which do not have a structured population, and accordingly only a single effective species. Our more general approach could allow for progress to be made investigating more heterogeneous interaction structures in and beyond ecology, such as other block-structured interaction matrices \cite{Kuczala_block_variances}, meta-population models on complex networks \cite{Grilli2015_metapopulaitons,Hanski2000_metapopulations,gravel2016stability}, or trophic levels \cite{Baron2020Diffusion,Johnson2014_TrophicCoherence,Baiser2013_foundationSpecies}.
    
\section*{Code availability}  

Codes can be found at \url{https://github.com/LylePoley/Cascade-Model.git}. A method for simulating the community dynamics with \cref{eq:GLV}, numerical solution procedures for solving the cascade model fixed point equations \cref{eq:FPeqnsCascade} and the data/code for producing all figures.

\section*{Acknowledgements}

We acknowledge funding from the Spanish Ministry of Science, Innovation and Universities, the Agency AEI and FEDER (EU) under the grant PACSS (RTI2018-093732-B-C22), the Maria de Maeztu program for Units of Excellence in R\&D (MDM-2017-0711) funded by MCIN/AEI/10.13039/501100011033, and the Engineering and Physical Sciences Research Council UK, grant number EP/T517823/1.


\onecolumngrid
\newpage

\setcounter{section}{0}
\setcounter{page}{1}
\setcounter{equation}{0}
\setcounter{figure}{0}
\setcounter{table}{0}

\renewcommand{\thesection}{S\arabic{section}} 		
\renewcommand{\thepage}{S\arabic{page}} 			
\renewcommand{\theequation}{S\arabic{equation}}  	
\renewcommand{\thefigure}{S\arabic{figure}}  		
\renewcommand{\thetable}{S\arabic{table}}  

\renewcommand*{\citenumfont}[1]{S#1}

\begin{center}
\textbf{\Large--- Supplemental Material ---}
\end{center}

\fontsize{12pt}{16}\selectfont
 
    \tableofcontents
    
    \newpage
    
    \section{Overview}
    This supplement contains further technical detail on the results in the main paper. In particular we report the generating functional analysis that was used to derive the effective dynamics and the subsequent stability analysis that was used deduce the phase diagrams in the main text.
    
    The document is structured as follows: 
    
    First, in \cref{appendix:pp_pairs} we derive the formula for the relation between the proportion of predator prey interaction pairs  $p$ in the community to the correlation parameter $\gamma$, $\gamma = \cos(\pi p)$. 
    
    Then, in \cref{appendix:GF} we outline the derivation of the effective single-species dynamics in Eq.~(6) of the main text. We then derive Eqs.~(13) in the main text, a set of self-consistent equations which provide the statistics of the species abundances, assuming a unique stable fixed-point. 
    
    In \cref{appendix:InfiniteSolution} we solve these fixed-point equations for the specific case of the Cascade Model, arriving at Eqs.~(26) of the main text. 
    
    In \cref{appendix:abundanceDistributions} we then use these results to derive expressions for hierarchy [Eq.~(34)], rank, and species abundance distributions [Eq.~(32)]. 
    
    In \cref{appendix:FPStability}, we then examine the stability of our system, detailing conditions for both linear instability and the point at which abundances diverge. We then compare our results against simulation, finding excellent agreement in \cref{appendix:numerical_verification_of_stability_curves}. We also demonstrate that the parameters $\nu$ and $\rho$ are stabilising and destabilising respectively by examining some informative special cases.
    
    Finally, in \cref{appendix:nalphaIndependence} we demonstrate that all fixed point properties of our system that are of interest to us are independent of $n(\alpha)$.

\section{Relation Between fraction of predator-prey pairs and \texorpdfstring{$\gamma$}{}, assuming Gaussian distributed interactions} \label{appendix:pp_pairs}
A pair of interaction coefficients $A^{ab}_{ij}, A^{ba}_{ji}$ are a predator-prey pair if their product is negative. We will use $X_1 = A^{ab}_{ij}$ and $X_2=A^{ba}_{ji}$ for the sake of this section and call $\mu_1 = \mu^{ab}, \mu_2 = \mu^{ba}, \sigma_1 = \sigma^{ab}, \sigma_2 = \sigma^{ba}, \gamma^{ab} = \gamma$ to avoid cluttering the argument with superfluous indices. 

The proportion of interaction pairs $X_1, X_2$ which are of predator prey type is then the sum of two  probabilities
\begin{align}
    \Prob(X_1X_2 < 0) = \Prob(X_1 < 0,~X_2 > 0) + \Prob(X_1 > 0,~X_2 < 0).
\end{align}
This can be evaluated using the Cholesky decomposition of $X_1, X_2$
\begin{align}
    X_1 &= \frac{\mu_1}{N} + \frac{\sigma_1}{\sqrt{N}} z_1, \nonumber \\
    X_2 &= \frac{\mu_2}{N} + \frac{\sigma_2}{\sqrt{N}} \lrb{\gamma z_1 + \sqrt{1 - \gamma^2} z_2},
\end{align}
where $z_1, z_2$ are uncorrelated, mean zero, unit variance random variables. We then have
\begin{align}
    \Prob(X_1 > 0,~X_2 < 0) &= \Prob\lrb{z_1 < -\frac{\mu_1}{\sigma_1}\frac{1}{\sqrt{N}},~z_1\gamma + z_2\sqrt{ 1 - \gamma^2} > -\frac{\mu_2}{\sigma_2}\frac{1}{\sqrt{N}}} \nonumber \\
    &= \Prob\lrb{z_1 < 0,~z_1\gamma + z_2\sqrt{ 1 - \gamma^2} > 0} + \mathcal{O}\lrb{\frac{1}{\sqrt{N}}},
\end{align}
where we have assumed that $\mu_{1, 2}$ and  $\sigma_{1, 2}$ are $\mathcal{O}(1)$. Therefore, to leading order in $N$, the probability we want is a region bounded by the lines $z_1 = 0$ and $\gamma z_1  = - \sqrt{1 - \gamma^2}~z_2$ in the $(z_1,z_2)$-plane. The joint distribution of $z_1, z_2$ is circularly symmetric as $z_1$ and $z_2$ are uncorrelated, hence the probability we want is proportional to the angle between these two lines, or
\begin{align}
    \Prob\lrb{z_1 < 0,~z_1\gamma + z_2\sqrt{ 1 - \gamma^2} > 0} = \frac{\arccos(\gamma)}{2\pi}.
\end{align}
A similar line of reasoning shows that $\Prob\lrb{X_1 > 0, X_2 < 0}$ has the same value, and so the proportion of predator-prey pairs $p$ is
\begin{align}
    p = \frac{\arccos(\gamma)}{\pi} + \mathcal{O}\lrb{\frac{1}{\sqrt{N}}}.
\end{align}
If $\gamma$ is not the same for all sub-communities, we find the proportion of predator-prey pairs by summing over all blocks
\begin{align}
    p = \frac{1}{\pi} \sum_{ab} n^a n^b \arccos(\gamma^{ab}).
\end{align}

\section{Derivation of The effective Dynamics}
\label{appendix:GF}
To find the effective process in Eq.~(6) in the main text, we follow \cite{BiolHandbookGF,Hertz_2016} and consider the MSRJD  \cite{De_Dominics_Peliti1978,De_Dominics1978,Janssen1976,MSR_formalism} functional integral of the GLV dynamics in Eq.~(1)
\begin{align}
	Z[h,\psi,A] =& \int D[x, \wh] 
	\exp \left(i \sum_{(a,i)}\int ds  \lrsqbracket{ 
			\wh^a_i(s) \lrb{
				\frac{\dot x_i^a(s)}{x_i^a(s)} - \lrsqbracket{ 
					1 + \sum_{(b,j)} A_{ij}^{ab} x_j^b(s) + h_i^a(s)}} } \right)\nonumber \\
	&\times \exp\left[i \sum_{(a,i)}\int ds\, x_i^a(s)\psi_i^a(s)\right]
	\label{app:eq:GenFunc},
\end{align}
where
\begin{align}
	D[x,\wh] &:= \prod_{(a, i)} \prod_{s = 0}^t \frac{dx_i^a(s) d\wh_i^a(s)}{\sqrt{2\pi}},
\end{align}
and where $(a, i)$ indicates a sum or product over both $a$ and $i$, so that, for example 
\begin{align}
    \sum_{(a, i)}x^a_i(t) \equiv \sum_{a = 1}^B\sum_{i = 1}^{N^a} x^a_i(t).
\end{align}
The functions $h^a_i(s)$ are perturbation fields, and the $\psi^a_i(s)$ are source terms. Taking functional derivatives of $Z[h, \psi, A]$ with respect to these functions generates moments of species abundances $x^a_i(s)$ as well as of the conjugate variables $\wh^a_i(s)$. For example we have
\begin{align}
    \funcdiff[Z]{h^a_i(s)} = -\avg{i\wh^a_i(s)}_Z, \hspace{1cm}
    \funcdiff[Z]{\psi^a_i(s)} = \avg{x^a_i(s)}_Z, \hspace{1cm}
    \funcddiff[Z]{\psi^a_i(s)}{h^a_i(s')} = -i\avg{x^a_i(s)\wh^a_i(s')}_Z. 
\end{align}
The notation $\avg{\dots}_Z$ denotes an average over the integral in \cref{app:eq:GenFunc}
\begin{multline}
	\avg{\dots}_Z \equiv \int D[x, \wh] (\dots)
	\exp \left(i \sum_{(a,i)}\int ds  \lrsqbracket{ 
			\wh^a_i(s) \lrb{
				\frac{\dot x_i^a(s)}{x_i^a(s)} - \lrsqbracket{ 
					1 + \sum_{(b,j)} A_{ij}^{ab} x_j^b(s) + h_i^a(s)}} } \right) \\
	\times \exp\left[i \sum_{(a,i)}\int ds\, x_i^a(s)\psi_i^a(s)\right],
\end{multline}

If $\psi^a_i(s) = 0$ and  $h^a_i(s) = 0$ in \cref{app:eq:GenFunc}, then the average $\avg{\dots}_Z$ constrains the system to follow the dynamics in Eq.~(1) in the main text. Further, if $\psi^a_i(s) = 0$ in \cref{app:eq:GenFunc}, then the average $\avg{\dots}_Z$ is normalised, so that $Z[h, \psi = 0, A] = 1$ \cite{De_Dominics1978}. 

We now average the generating functional $Z$ over realisations of the interaction matrix $A^{ab}_{ij}$, and note that such a procedure can also be performed using the cavity method \cite{BuninOpenLVDynamics,Barbier2156, roy2019numerical}.

\subsection{Disorder average}
Denoting averages over realisations of $A$ by an overbar, we find
\begin{multline}
    \overline{\explrbr{- i \sum_{(a, i),(b, j)}\int ds\ A_{ij}^{ab}\ \wh_i^a(s) x_j^b(s)}} = \ 1 - i\sum_{(a, i),(b, j)}\int ds\ \overline{A_{ij}^{ab}}\ \wh_i^a(s) x_j^b(s) \\
    - \sum_{(a, i),(b, j)}\sum_{(c, k),(d, l)}\int ds ds'\ \overline{A_{ij}^{ab}A_{kl}^{cd}}\ \wh_i^a(s) x_j^b(s)\wh_k^c(s') x_l^d(s') + \cdots.
\end{multline}
To leading order in $N$, the averaged functional $\overline{Z[h,\psi,A]}$ only depends on the first two moments of $A$, provided higher moments drop sufficiently fast with growing $N$. Upon averaging the above and re-exponentiating we find
\begin{multline}
    \overline{\explrbr{- i \sum_{(a, i),(b, j)}\int ds\ A_{ij}^{ab}\ \wh_i^a(s) x_j^b(s)}} = 	\explrbr{- N \sum_{a, b} n^an^b \int ds\  \mu^{ab}\  P^a(s) M^b(s)} \\
	\hfill \explrbr{-  N \sum_{a, b}n^an^b\int dsds'\ \lrsb{\half(\sigma^{ab})^2\  L^a(s, s') C^b(s,s')
	+ \gamma^{ab}  \sigma^{ab} \sigma^{ba}\ K^a(s,s') K^b(s',s)} + \mathcal{O}(N^0)},
\end{multline}
where $n^a \equiv N^a/N$, and where we have introduced the following macroscopic parameters as the level of a single sub-community 
\begin{align}
\begin{aligned}
	M^a(t) &:= \frac{1}{N^a} \sum_{i = 1}^{N^a} x^a_i(t), \\
	P^a(t) &:= \frac{i}{N^a} \sum_{i = 1}^{N^a} \wh^a_i(t), \\
	C^a(t,t') &:= \frac{1}{N^a} \sum_{i = 1}^{N^a} x^a_i(t)x^a_i(t'), \\
	L^a(t,t') &:= \frac{1}{N^a} \sum_{i = 1}^{N^a} \wh^a_i(t) \wh^a_i(t'), \\
	K^a(t,t') &:= \frac{1}{N^a} \sum_{i = 1}^{N^a} \wh^a_i(t) x^a_i(t'). 
\end{aligned}
\end{align}
We impose these definitions in the generating functional $Z$ using Dirac delta functions in their complex exponential representation, e.g.
\begin{align}
	1 &= \int \prod_{a = 1}^B \prod_{s} d
	M^a(s)\ \delta\lrb{M^a(s) - \frac{1}{N^a}\sum_i^{N^a}x^a_i(s)} \nonumber
	\\
	&\equiv \int D\{M, \wh[M]\}
	\explrbr{i \sum_a N^a \int ds\  \wh[M]^a(s) \lrsqbracket{M^a(s) - \frac{1}{N^a} \sum_i^{N^a} x^a_i(s)}}.
\end{align}
We absorb prefactors of $2\pi$ and $N$ that result from this procedure into the measure $D\{M, \wh[M]\}$. In the saddle-point approximation, this prefactor has no bearing on our results.

After averaging over $A$ and insertion of the order parameters into $\overline{Z[h, \psi, A]}$ we have 
\begin{align}
	\overline{Z[h,\psi, A]} = \int D\{M,\wh[M]\}\cdots D\{K, \wh[K]\} \explrbr{N\lrsb{\Psi[M,\wh[M],\cdots,K, \wh[K]] + \Phi[M,\cdots,K] + \Omega[\wh[M],\cdots,\wh[K]]}} \label{app:eq:GenFuncAvg},
\end{align}
where
\begin{align}
&\begin{aligned}
	\Psi =&\ i \sum_a n^a \int ds\ \lrb{M^a(s)\wh[M]^a(s) + P^a(s)\wh[P]^a(s)} \\
	 &+ i \sum_a n^a \int dsds'\ \lrb{L^a(s,s')\wh[L]^a(s,s') + C^a(s,s')\wh[C]^a(s,s') + K^a(s,s')\wh[K]^a(s,s')}, 
\end{aligned}\\[0.4cm]
&\begin{aligned}
	\Phi =&\ - 
	\sum_{a,b} n^a n^b \int ds\ \lrb{ \mu^{ab} P^a(s) M^b(s)} \\
	& - \sum_{a,b} n^a n^b \int dsds'\ \lrb{ \half (\sigma^{ab})^2 L^a(s,s')C^b(s,s') + \gamma^{ab} \sigma^{ab}\sigma^{ba} K^a(s,s') K^b(s',s)},  
\end{aligned}\\[0.4cm]
&\begin{aligned}
	\Omega = \ln \int D[x,\wh] 
    &\explrbr{ i\sum_{a} \int ds\ \lrsb{\wh^a(s) \lrsqbracket{\frac{\dot x^a(s)}{x^a(s)} - \lrbracket{1 - x^a(s) + h^a(s)}} + \psi^a(s) x^a(s)}} \\
	&\hspace{-3em}\times\explrbr{- i \sum_{a} \int ds\ \lrsb{\wh[M]^a(s)x^a(s) + \wh[P]^a(s)i\wh^a(s)}} \\
	&\hspace{-3em}\times\explrbr{- i \sum_{a} \int dsds'\ \lrsb{ \wh[L]^a(s,s')\wh^a(s) \wh^a(s') + \wh[C]^a(s,s')x^a(s) x^a(s') + \wh[K]^a(s,s') \wh^a(s) x^a(s')}}.
\end{aligned}
\end{align}
The common factor of $N$ in the exponent in Eq.~(\ref{app:eq:GenFuncAvg}) makes the integral amenable to a saddle-point approximation for large $N$. At the saddle point, we find that the macroscopic hatted and un-hatted parameters satisfy 
\begin{align}
\begin{aligned}
	i\wh[M]^a(t) &= \sum_{b = 1}^B n^b s^b(t) \mu^{ba}, \\
	i\wh[P]^a(t) &= \sum_{b = 1}^B \mu^{ab} n^b M^b(t), \\
	i\wh[C]^a(t,t') &= \half \sum_{b = 1}^B n^b L^b(t,t')(\sigma^{ba})^2,  \\
	i\wh[L]^a(t,t') &= \half \sum_{b = 1}^B (\sigma^{ab})^2 n^b C^b(t,t'), \\
	i\wh[K]^a(t,t') &= \half \sum_{b = 1}^B \gamma^{ab} \sigma^{ab} \sigma^{ba} n^b K^b(t', t).
\end{aligned}
&&
\begin{aligned}
	M^a(t) &= \frac{1}{N^a} \sum_{i = 1}^{N^a} \overline{\avg{x^a_i(t)}_Z}, \\
	P^a(t) &= \frac{i}{N^a} \sum_{i = 1}^{N^a} \overline{\avg{\wh^a_i(t)}_Z}, \\
	C^a(t, t') &= \frac{1}{N^a} \sum_{i = 1}^{N^a} \overline{\avg{x^a_i(t)x^a_i(t')}_Z}, \\
	L^a(t, t') &= \frac{1}{N^a} \sum_{i = 1}^{N^a} \overline{\avg{\wh^a_i(t)\wh^a_i(t')}_Z}, \\
	K^a(t, t') &= \frac{1}{N^a} \sum_{i = 1}^{N^a} \overline{\avg{\wh^a_i(t)x^a_i(t')}_Z}. \label{app:eq:saddle}	
\end{aligned} 
\end{align}
In particular, we see that $P^a(s)$ and $L^a(s, s')$ vanish at the saddle point as
\begin{align}
    \avg{\wh^a_i(t)}_Z = \funcdiff[{Z[h, \psi = 0, A]}]{h^a_i(s)} = \funcdiff[(1)]{h^a_i(s)} = 0, \label{app:eq:Pvanishes}
\end{align}
and similarly for $\avg{\wh^a_i(t) \wh^a_i(t')}_Z$ (see e.g. \cite{Galla_2018,BiolHandbookGF} for similar calculations).

\subsection{Effective Dynamics}
For the rest of the calculation we set $\psi^a_i(s) = h^a_i(s) = 0$, they are fields introduced with the sole purpose of producing correlation functions such as that in \cref{app:eq:Pvanishes} and are no longer needed. Substituting the saddle-point conditions in \cref{app:eq:saddle} into the disorder-averaged generating functional in \cref{app:eq:GenFuncAvg}, we obtain a generating functional that factorises as follows
\begin{align}
	\overline{Z[h = 0, \psi = 0, A]} = \prod_{a = 1}^{B} \lrb{Z_{\text{eff}}^a}^{N^a},
\end{align}
where $Z_{\text{eff}}^a$ is an effective generating functional for the $a^{\text{th}}$ block
\begin{align}
	Z^a_{\text{eff}} &= \int \prod_s \frac{dx^a(s) d\wh^a(s)}{\sqrt{2\pi}} \nonumber \\
	&\times \explrbr{ i \int ds\ \wh^a(s) \lrsqbracket{\frac{\dot x^a(s)}{x^a(s)} - \lrbracket{1 - x^a(s)}}}, \nonumber\\
	&\times  \explrbr{ - i \int ds\ \wh^a(s) \lrsqbracket{\sum_b n^b\lrb{\mu^{ab} M^b(s) + \gamma^{ab} \sigma^{ab}\sigma^{ba} \int ds'\ x^a(s')G^b(s,s')}}}, \nonumber\\
	&\times  \explrbr{-  i\int dsds'\ \lrsb{ \half\sum_{b} n^b (\sigma^{ab})^2 C^b(s,s')\wh^a(s) \wh^a(s')}},
\end{align}
where we define $G^a(t,t') = -iK^a(t', t)$. 

The object $Z^a_\text{eff}$ is recognised as the generating functional of the process in Eq.~(6) of the main text
\begin{align}
	\dot x^a(t) &= x^a(t)\lrb{1 - x^a(t) + \sum_{b = 1}^B n^b \lrsb{\mu^{ab}M^b(t) + \gamma^{ab}\sigma^{ab}\sigma^{ba}\int_0^t dt'\ x^a(t')G^b(t, t')} + \eta^a(t)}, \label{app:eq:effectiveDynamics}
\end{align}
with $M^a(t), C^a(t,t'), G^a(t,t')$ the average abundance, correlation, and response functions for abundances in block $a$, 
\begin{gather}
	M^a(t) := \avg{x^a(t)}, \hspace{0.5cm}
	C^a(t,t') := \avg{x^a(t)x^a(t')}, \hspace{0.5cm}
	G^a(t, t') := \funcdiff[\avg{x^a(t)}]{\eta^a(t')}, \label{app:eq:MCandG}
\end{gather}
and $\eta^a(t)$ colored Gaussian noise 
\begin{gather}
	\avg{\eta^a(t)} = 0,\hspace{1cm}
	\avg{\eta^a(t)\eta^a(t')} = \sum_{b = 1}^B n^b (\sigma^{ab})^2 C^a(t,t'), \label{app:eq:eta}
\end{gather}
where $\avg{\dots}$ is an average over the effective process in \cref{app:eq:effectiveDynamics}.  
\subsection{Fixed Point Equations}\label{appendix:fixedpoint}
We now suppose that the system \cref{app:eq:effectiveDynamics} reaches a fixed point such that $\dot x^a(t) \to 0, x^a(t) \to x^a_*$ as $t \to \infty$. Under this assumption, we replace $M^a(t) \to M^a, C^a(t, t') \to q^a, \eta^a(t) \to \sqrt{\sum_b (\sigma^{ab})^2 n^b q^b}z^a$, where the $z^a$ are independent mean-zero static Gaussian random variables with unit variance. Further assuming that the response function $G^a(t, t')$ is a function only of time differences $\tau = t - t'$, we deduce that 
\begin{align}
    0 &= x^a_*\lrb{1 - x^a_* + \sum_b n^b \lrsb{\mu^{ab}M^b + \gamma^{ab}\sigma^{ab}\sigma^{ba} x^a_* \chi^a} + \sqrt{\sum_b (\sigma^{ab})^2 n^b q^b}z^a},
\end{align}
where we have defined the integrated response function $\chi^a = \int_0^\infty d\tau G^a(\tau)$. 

Similarly to cases with no block structure \cite{Galla_2018,BuninOpenLVDynamics}, we assume that $1 - \sum_b\gamma^{ab}\sigma^{ab}\sigma^{ba}n^b\chi^b > 0$ and find
    \begin{align}\label{app:eq:fpsolution}
    	x^a_* = \max\lrb{0, \frac{1 + \sum_b \mu^{ab} n^b M^b + \sqrt{\sum_b (\sigma^{ab})^2 n^b q^b}z^a}{1 - \sum_b \gamma^{ab} \sigma^{ab} \sigma^{ba} n^b \chi^b}},
    \end{align}
    so that $x^a_* \geq 0$, and has a non-zero value when the non-zero fixed point is positive.
    
    Following \cite{1overf_noise,Galla_2018}, we can self-consistently determine $\chi^a, M^a, q^a$ from \cref{app:eq:fpsolution} together with \cref{app:eq:effectiveDynamics,app:eq:MCandG,app:eq:eta}. One obtains
    \begin{align}
        \chi^a &= \frac{1}{1 - \sum_b\gamma^{ab}\sigma^{ab}\sigma^{ba}n^b\chi^b}w_0(\Delta^a) , \nonumber \\
        M^a &= \frac{\sqrt{\sum_b(\sigma^{ab})^2n^bq^b}}{1 - \sum_b\gamma^{ab}\sigma^{ab}\sigma^{ba}n^b\chi^b}w_1(\Delta^a), \nonumber \\
        q^a &= \frac{\sum_b(\sigma^{ab})^2n^bq^b}{\lrb{1 - \sum_b\gamma^{ab}\sigma^{ab}\sigma^{ba}n^b\chi^b}^2}w_2(\Delta^a),
        \label[pluralequation]{app:eq:FPeqnsGeneral}
    \end{align}
    where
    \begin{align}
        \Delta^a &\equiv \frac{1 + \sum_b\mu^{ab}n^bM^b}{\sqrt{\sum_b(\sigma^{ab})^2n^bq^b}}, \\
        w_k(\Delta^a) &\equiv \int_{-\infty}^{\Delta^a} \frac{dz}{\sqrt{2\pi}}~e^{-\half z^2}\lrb{\Delta^a - z}^k. \label{app:eq:wdefn}
    \end{align}
    Hence we arrive at Eqs.~(13) in the main text.
\section{Derivation of Fixed point Eqs.~(26)}
\label{appendix:InfiniteSolution}
There are a number of steps to the following derivation, the results of which allow us to calculate the average abundance $\E{M}$ and survival rate $\E{\phi}$ in the $B \to \infty$ limit of the cascade model. 

First, we take the $B \to \infty$ limit of \cref{app:eq:FPeqnsGeneral} to get \cref{app:eq:FPeqnsGeneralBinfty}. Then we impose cascade model interactions, deriving Eqs.~\eqref{app:eq:FP1} through \eqref{app:eq:FP5}.

Once we have found the set of equations to be solved we set about solving them. We first set up Eqs.~(21) in \cref{appendix:uRelationsDerivation} as self consistent integrals over the species index $\alpha$. We then change variables from $\alpha$ to $\Delta$ in \cref{appendix:changeOfVariables} to make these integrals explicit. The result of this change in variables is Eqs.~(26), equations for determining $u, \Delta_0, \Delta_1$ as functions of $\nu, \sigma, \rho, \gamma$.

Finally, we derive expressions for $\E{M}$ and $\E{\phi}$ as explicit functions of $\mu, \nu, \sigma, \rho, \gamma$ and $u, \Delta_0, \Delta_1$.
\subsection{Taking the \texorpdfstring{$B \to \infty$}{B to infinity} Limit of Eqs.~(13) in the main text}
\label{appendix:btoinfty}
We will take for example the term $\sum_{b = 1}^B \mu^{ab} n^b M^b$, which appears in the second of Eqs.~(14). The same steps apply analogously to all similar terms in Eqs.~(13) and (14). Writing $\alpha=a/B$, $\beta=b/B$ as well as $n(\beta) := Bn^b$, we have
\begin{align}
    \lim_{B \to \infty} \sum_{b = 1}^B \mu^{ab} n^b M^b  = \int_0^1 d\beta\ \mu(\alpha, \beta) n(\beta) M(\beta),
\end{align}
where we have introduced
\begin{align}
    \mu(\alpha, \beta) \equiv \mu^{ab}, \hspace{0.5cm} 
    M(\beta) \equiv M^b.
\end{align}

\noindent The condition $\sum_b n^b = 1$ becomes an integral as well
\begin{align}
    \lim_{B \to \infty} \sum_b n^b  = \int_0^1 d\beta~ n(\beta) = 1. \label{app:eq:normalisationofn}
\end{align}
In the $B \to \infty$ limit our general fixed point equations [see Eqs.~(13)] are therefore
\begin{align}
    u(\alpha) &= 1 - \int_0^1 d\beta~ \gamma(\alpha, \beta)\sigma(\alpha, \beta)\sigma(\beta, \alpha)n(\beta)\chi(\beta),\nonumber\\
    \chi(\alpha)u(\alpha) &= w_0(\Delta(\alpha)),\nonumber \\
    M(\alpha)u(\alpha) &= w_1(\Delta(\alpha)) \sqrt{\int_0^1 d\beta~ \sigma(\alpha, \beta)^2n(\beta)q(\beta)}, \nonumber\\
    q(\alpha)u(\alpha)^2 &= w_2(\Delta(\alpha)) \int_0^1 d\beta~ \sigma(\alpha, \beta)^2n(\beta)q(\beta), \nonumber\\
    \Delta(\alpha) &= \frac{1 + \int_0^1 d\beta~ \mu(\alpha, \beta)n(\beta)M(\beta)}{\sqrt{\int_0^1 d\beta~ \sigma(\alpha, \beta)^2n(\beta)q(\beta)}}
    \label[pluralequation]{app:eq:FPeqnsGeneralBinfty}
\end{align}
with
\begin{align}
    \mu\lrb{\frac aB, \frac bB} = \mu^{ab}, \hspace{1cm}
    \sigma\lrb{\frac aB, \frac bB} = \sigma^{ab}, \hspace{1cm}
    \gamma\lrb{\frac aB, \frac bB} = \gamma^{ab}.
\end{align}
In the cascade model we choose the matrices $\mu^{ab}, \sigma^{ab}, \gamma^{ab}$ as in Eq.~(4) in the main paper. For the matrix $\mu^{ab}$, this leads to
\begin{align}
    \mu(\alpha, \beta) = 
    \begin{cases}
		\mu - \nu,\ &\alpha < \beta \\
		\mu(\alpha),\ &\alpha = \beta \\
		\mu + \nu,\ &\alpha > \beta
	\end{cases},
\end{align}
The $\alpha = \beta$ contribution is negligible in the limit $B\to \infty$. We therefore conclude
\begin{align}
    \lim_{B \to \infty} \sum_b \mu^{ab} n^b M^b = \mu\E{M} + \nu\lrb{\int_0^\alpha d\beta\ n(\beta)M(\beta) - \int_\alpha^1 d\beta\ n(\beta)M(\beta)}.
\end{align}
The notation $\E{\dots}$ indicates an average over the index $\alpha$, that is  
\be\label{eq:square_bracket}
\E{\dots} = \int_0^1d\alpha~n(\alpha)(\dots).
\ee
Similarly, one finds
\begin{align}
    \lim_{B \to \infty} \sum_b \lrb{\sigma^{ab}}^2 n^b q^b = \sigma^2\lrb{\rho^2\int_0^\alpha d\beta\ n(\beta)q(\beta) + \frac{1}{\rho^2} \int_\alpha^1 d\beta\ n(\beta)q(\beta)}.
\end{align}
Inspired by the similarity in form of the above, we define two operators with action on a function $f(\alpha)$,
\begin{align}
     f_\nu(\alpha) & \equiv \nu \int_0^\alpha d\beta\ n(\beta)f(\beta) - \nu \int_\alpha^1 d\beta\ n(\beta)f(\beta), \nonumber \\
     f_\rho(\alpha) & \equiv \rho^2\int_0^\alpha d\beta\ n(\beta)f(\beta) + \frac{1}{\rho^2} \int_\alpha^1 d\beta\ n(\beta)f(\beta). \label[pluralequation]{app:eq:f_mu_sigma}
\end{align}
Eqs.~(13) may now be written as
\begin{align}
    u &= 1 - \gamma \sigma^2\E{\chi}, \label{app:eq:FP1} \\
	\chi(\alpha) u &= w_0[\Delta(\alpha)], \label{app:eq:FP2} \\
	M(\alpha) u &= w_1[\Delta(\alpha)] \sigma \sqrt{q_\rho(\alpha)}, \label{app:eq:FP3}\\
	q(\alpha) u^2 &= w_2[\Delta(\alpha)] \sigma^2 q_\rho(\alpha), \label{app:eq:FP4}\\
	\Delta(\alpha) &= \frac{1 + \mu\E{M} + M_\nu(\alpha)}{\sigma \sqrt{q_\rho(\alpha)}}. \label{app:eq:FP5}
\end{align}
A specific use of the average notation $\E{\dots}$ we will use frequently is the average of a composite function $f[\Delta(\alpha)]$, where $f$ is again any function. We write
\begin{align}\label{app:eq:squarebracketaverage}
    \E{f(\Delta)} \equiv \int_0^1d\alpha~n(\alpha)f[\Delta(\alpha)].
\end{align}
\subsection{Derivation of Eqs.~(21)} \label{appendix:uRelationsDerivation}
To derive the first of Eqs.~(21), we first average \cref{app:eq:FP2} to obtain
\begin{align}
    \E{\chi}u &= \E{w_0(\Delta)},
\end{align}
Now, substituting this result into \cref{app:eq:FP1} gives
\begin{align}
    u^2 = u - \gamma\sigma^2\E{w_0(\Delta)}. \label{app:eq:FPsol1}
\end{align}
 \cref{app:eq:FPsol1} is the first equation in Eqs.~(26).

To derive the second of Eqs.~(21), we differentiate $q_\rho(\alpha)$ with respect to $\alpha$, finding
\begin{align}
    \diff{\alpha}q_\rho(\alpha) &= \lrb{\rho^2 - \rho^{-2}}n(\alpha)q(\alpha), \nonumber \\
    &= \frac{\ell\sigma^2\ln\rho^2}{u^2}n(\alpha)w_2(\Delta(\alpha))q_\rho(\alpha), \label{app:eq:q_rhoODE}
\end{align}
where the second line follows from use of \cref{app:eq:FP4}, and where $\ell$ is the logarithmic mean of $\rho^2$ and $\rho^{-2}$, that is 
\begin{align}
    \ell \equiv \Lm\lrb{\rho^2, \frac{1}{\rho^2}}, \label{app:eq:ell}
\end{align}
with
\begin{align}
    \Lm\lrb{x, y} \equiv \frac{x - y}{\ln x - \ln y}.
\end{align}
\cref{app:eq:q_rhoODE} is a linear differential equation for $q_\rho(\alpha)$, with boundaries [from \cref{app:eq:f_mu_sigma}]
\begin{align}
    q_\rho(0) &= \E{q}\rho^{-2}, \\
    q_\rho(1) &= \E{q}\rho^2, \label{app:eq:qrho}
\end{align}
we note that either boundary can be used as an initial or final condition to solve \cref{app:eq:q_rhoODE} and the other will automatically be satisfied.
The solution is
\begin{align}
    q_\rho(\alpha) = A\explrbr{\frac{\ell\sigma^2  \ln\rho^2}{u^2}\int_0^\alpha d\beta~n(\beta)w_2(\beta)}, \label{app:eq:Cq}
\end{align}
for some constant $A$. Substituting in the boundary conditions gives
\begin{align}
    A &= \rho^{-2}\E{q}, \\
    u^2 &= \ell\sigma^2\E{w_2(\Delta)}.  \label{app:eq:usig}
\end{align}
\subsection{An implicit expression for \texorpdfstring{$\Delta(\alpha)$}{Delta}}
So far, we have the following three relations [see \cref{app:eq:FPsol1,app:eq:usig,app:eq:normalisationofn}]
\begin{align}
    u(1 - u) &= \gamma \sigma^2 \E{w_0(\Delta)}, \nonumber\\
    u^2 &= \ell\sigma^2 \E{w_2(\Delta)}, \nonumber\\
    1 &= \int_0^1d\alpha~n(\alpha). \label[pluralequation]{app:eq:FPeqnsCascadealpha}
\end{align}
One obtains Eqs.~(26), the equations necessary for obtaining $u, \Delta_0, \Delta_1$ from our model parameters, from the above by changing variables from $\alpha$ to $\Delta$ in the three integrals, $\E{w_0(\Delta)}, \E{w_2(\Delta)}$ and $\int_0^1d\alpha~n(\alpha)$. In order to find this change of variables, we will first derive Eq.~(23) in the main text, that is, we find a differential equation linking $\Delta(\alpha)$ to $\alpha$. To do this, we find an expression for $M_\nu(\alpha)$, which will give us an implicit expression for $\Delta(\alpha)$ by \cref{app:eq:FP5}, which in turn yields Eq.~(23) when differentiated.

First, we substitute \cref{app:eq:FP3} into \cref{app:eq:FP5} to get
\begin{align}
    M(\alpha)u = \frac{w_1(\Delta(\alpha))}{\Delta(\alpha)}\lrb{1+\mu\E{M} + M_\nu(\alpha)}. \label{app:eq:malpha1}
\end{align}
We can find $1 + \mu\E{M} + M_\nu(\alpha)$ in a similar manner to our derivation of $q_\sigma(\alpha)$. We have
\begin{align}
    \diff[]{\alpha}\lrb{1 + \mu\E{M} + M_\nu(\alpha)} &= 2\nu n(\alpha)M(\alpha), \nonumber \\
    &= \frac{2\nu}{u}n(\alpha)\frac{w_1(\Delta(\alpha))}{\Delta(\alpha)}\lrb{1 + \mu\E{M} + M_\nu(\alpha)}.
\end{align}
This is a differential equation for $1 + \mu\E{M} + M_\nu(\alpha)$, with boundaries [see \cref{app:eq:f_mu_sigma}]
\begin{align}
    1 + \mu\E{M} + M_\nu(0) &= 1 + (\mu-\nu)\E{M}, \\
    1 + \mu\E{M} + M_\nu(1) &= 1 + (\mu+\nu)\E{M}.
\end{align}
The solution is therefore
\begin{align}
    1 + \mu\E{M} + M_\nu(\alpha) &= A \explrbr{\frac{2\nu}{u}\int_0^\alpha d\beta~n(\beta)\frac{w_1(\Delta(\beta))}{\Delta(\beta)}}, \label{app:eq:CM}
\end{align}
where the boundary conditions give
\begin{align}
    A &= 1 + (\mu - \nu)\E{M}, \\
    u &= \Lmbr{\frac{1}{\E{M}} + \mu + \nu, \frac{1}{\E{M}} + \mu - \nu}\E{\frac{w_1(\Delta)}{\Delta}},  \label{app:eq:umu}
\end{align}
where we write $\E{w_1(\Delta)/\Delta} \equiv \int_0^1d\alpha~n(\alpha)w_1(\Delta(\alpha))/\Delta(\alpha)$. 

On substituting \cref{app:eq:Cq,app:eq:CM} into \cref{app:eq:FP5} we now have an expression for $\Delta(\alpha)$
\begin{align}
	\Delta(\alpha) = \Delta_0\explrbr{\frac{2\nu}{u}\int_0^\alpha d\beta~n(\beta)\frac{ w_1(\Delta(\beta))}{\Delta(\beta)} 
	-\frac{\sigma^2\ell\ln\rho}{u^2}\int_0^\alpha d\beta~ n(\beta)w_2(\Delta(\beta))}, \label{app:eq:deltaexp}
\end{align}
\subsection{Derivation of Eq.~(23) and Eqs.~(26)}
\label{appendix:changeOfVariables}
We may now differentiate both sides of \cref{app:eq:deltaexp} with respect to $\alpha$ to get Eq.~(23) in the main text
\begin{align}
    \diff[\Delta]{\alpha} &= \frac{n(\alpha)}{\theta(\Delta, u)}, \label{app:eq:DeltaODE}
\end{align}
where
\begin{align}
    \theta(\Delta, u) \equiv \frac{u^2}{2}\frac{1}{u\nu~w_1(\Delta) 
	-\ell\sigma^2\ln\rho~\Delta w_2(\Delta)}.
	\label{app:eq:thetadefn}
\end{align}
We are now in a position to change variables from $\alpha$ to $\Delta$ in \cref{app:eq:FPeqnsCascadealpha}. Using \cref{app:eq:DeltaODE} we can write 
\begin{align}
    \E{w_0(\Delta)} = \int_0^1d\alpha~n(\alpha)w_0(\Delta(\alpha)) = \int_{\Delta_0}^{\Delta_1}d\Delta~\theta(\Delta, u)w_0(\Delta),
\end{align}
and similarly for $\E{w_2(\Delta)}$ and $\E{w_1(\Delta)/\Delta}$. The result is Eqs.~(26) in the main text
\begin{align}
    u(1 - u) &= \gamma\sigma^2 \int_{\Delta_0}^{\Delta_1}d\Delta~ \theta(\Delta, u)w_0(\Delta), \nonumber\\
    u^2 &=  \ell\sigma^2 \int_{\Delta_0}^{\Delta_1}d\Delta~ \theta(\Delta, u)w_2(\Delta), \nonumber \\
    1 &= \int_{\Delta_0}^{\Delta_1}d\Delta~ \theta(\Delta, u). \label[pluralequation]{app:eq:FPeqnsCascade}
\end{align}
Hence, given the parameters $\nu, \sigma, \rho, \gamma$, \cref{app:eq:FPeqnsCascade} can be solved to find $u, \Delta_0$ and $\Delta_1$. Therefore we can readily obtain $u, \Delta_0$ and $\Delta_1$ as known functions of our system parameters, i.e.
\begin{align}
    u &\equiv u(\nu, \sigma, \rho, \gamma), \\
    \Delta_0 &\equiv \Delta_0(\nu, \sigma, \rho, \gamma), \\
    \Delta_1 &\equiv \Delta_1(\nu, \sigma, \rho, \gamma).
\end{align}
\subsection{Expressions for \texorpdfstring{$\E{M}, \E{\phi}$}{} [derivation of Eqs.~(27) and (28)]}
\subsubsection{Expression for \texorpdfstring{$\E{M}$}{}}
To find an expression for $\E{M}$, we evaluate \cref{app:eq:FP5} at $\alpha = 0$ and $\alpha = 1$, giving
\begin{align}
    \Delta_0 &= \frac{1 + (\mu - \nu)\E{M}}{\lrb{\sigma/\rho}\sqrt{\E{q}}}, \label{app:eq:Delta0}\\
    \Delta_1 &= \frac{1 + (\mu + \nu)\E{M}}{\sigma\rho~\sqrt{\E{q}}}. \label{app:eq:Delta1}
\end{align}
Alternatively, one can evaluate \cref{app:eq:deltaexp} at $\alpha = 0$ and $\alpha = 1$ and use \cref{app:eq:umu,app:eq:usig} to obtain the same expressions. Taking the ratio of the above expressions, we obtain
\begin{align}
    \frac{\Delta_1\rho^2}{\Delta_0} = \frac{1 + (\mu + \nu)\E{M}}{1 + (\mu - \nu)\E{M}}. \label{app:eq:Mratio}
\end{align}
Re-arranging for $\E{M}$ then gives Eq.~(27) in the main text:
\begin{align}
    \frac{1}{\E{M}} + \mu = \frac{\Delta_1\rho^2 + \Delta_0}{\Delta_1\rho^2 - \Delta_0}\nu. \label{app:eq:M}
\end{align}
\subsubsection{Expression for \texorpdfstring{$\E{\phi}$}{}}
Eq.~(28) follows from recognising that $\E{\phi} = \E{w_0(\Delta)}$ [since $\phi(\alpha) = w_0(\Delta(\alpha))$]. Therefore, by the first equation in \cref{app:eq:FPeqnsCascade} [or the first equation in \cref{app:eq:FPeqnsCascadealpha}, or \cref{app:eq:FPsol1}] we find Eq.~(28) of the main text
\begin{align}
    \E{\phi} = \frac{u(1-u)}{\gamma \sigma^2}. \label{app:eq:phi}
\end{align}

\subsubsection{The \texorpdfstring{$\nu \to 0$}{} limit of Eq.~(27) and the \texorpdfstring{$\gamma \to 0$}{} and \texorpdfstring{$\sigma \to 0$}{} limits of Eq.~(28)}
\label{appendix:MandPhiLimits}
As mentioned in the main text, we have to be careful with Eqs.~(27) and (28) [equivalently, \cref{app:eq:M,app:eq:phi}] if either of $\nu$ or $\gamma$ is set to zero. Similar care must be taken when $\sigma$ is set to zero. However, this would imply that there is no disorder in our system, and we are only interested in disordered interactions in this work, therefore we assume $\sigma > 0$. 

To find the $\nu \to 0$ limit of \cref{app:eq:M} we first note that if $\nu = 0$, then by \cref{app:eq:Mratio} $\Delta_1\rho^2 = \Delta_0$, so that the right hand side of \cref{app:eq:M} is in an indeterminate form. To find the appropriate limit we use \cref{app:eq:umu}, which we repeat here
\begin{align}
    u = \Lm\lrb{\frac{1}{\E{M}} + \mu - \nu, \frac{1}{\E{M}} + \mu + \nu}\E{\frac{w_1(\Delta)}{\Delta}}.
\end{align}
In principle, the above applies for all $\nu, \sigma, \rho, \gamma$ and therefore has greater applicability than \cref{app:eq:M}. However, when it does apply, \cref{app:eq:M} is simpler, less computationally expensive, and makes clearer the relationship between $\E{M}$ and $\mu$. It is easy to check\ that $\Lm(x, x) = x$ for any $x$, and so if $\nu = 0$ the above becomes
\begin{align}
    u = \lrb{\frac{1}{\E{M}} + \mu}\E{\frac{w_1(\Delta)}{\Delta}},
\end{align}
from which $\E{M}$ is readily obtained.

To find the $\gamma \to 0$ limit of \cref{app:eq:phi} we must look back at \cref{app:eq:FPeqnsCascade}, if we rearrange the first two of these equations for $u$ we find
\begin{align}
    u = \frac{\E{w_2(\Delta)}}{\E{w_2(\Delta)} + \gamma/\ell \E{w_0(\Delta)}}.
\end{align}
Hence, if $\gamma = 0$ then $u = 1$. Now we can simply solve the first and last of \cref{app:eq:FPeqnsCascade} for $\Delta_0, \Delta_1$
\begin{align}
    1 &= \ell\sigma^2\int_{\Delta_0}^{\Delta_1} d\Delta~\theta(\Delta, u = 1)w_2(\Delta), \\
    1 &= \int_{\Delta_0}^{\Delta_1} d\Delta~\theta(\Delta, u = 1).
\end{align}
With the values of $\Delta_0$ and $\Delta_1$ known, one then has two methods for computing $\E{\phi} = \E{w_0(\Delta)}$. We can directly compute it as an integral over $\Delta$
\begin{align}
    \E{\phi} = \int_{\Delta_0}^{\Delta_1} d\Delta~\theta(\Delta, u = 1)w_0(\Delta),
\end{align}
or we can solve the differential equation \cref{app:eq:DeltaODE} for the whole function $\Delta(\alpha)$, and compute
\begin{align}
    \E{\phi} = \int_0^1 d\alpha~n(\alpha)w_0(\Delta(\alpha)).
\end{align}
%
\section{Abundance Distributions}\label{appendix:abundanceDistributions}
\subsection{Species abundance distributions (SADs), derivation of Eq.~(32)}
We wish to find the probability that the abundance in sub-community $a$ is $x$, $\Prob(x|a)$. Setting $x^a_*(\eta^a_*) = x$ in Eq.~(12) of the main text gives
\begin{align}\label{app:eq:aux1}
    x = \max\lrb{0, \frac{1 + \sum_b \mu^{ab} n^b M^b + \sqrt{\sum_b (\sigma^{ab})^2n^b q^b} z^a}{1 - \sum_b \gamma^{ab} \sigma^{ab} \sigma^{ba} n^b \chi^b}},
\end{align}
where we have written $\eta^a_* = \sqrt{\sum_b (\sigma^{ab})^2n^b q^b} z^a$ for a mean-zero, unit variance Gaussian random variable $z^a$. 

\cref{app:eq:aux1} implies that the distribution of $x^a_*$ has two contributions. First, a delta function at $x = 0$ with a weight equal to the probability of that species going extinct (having zero abundance). Second, a Gaussian distribution with mean $\frac{1 + \sum_b \mu^{ab}n^bM^b}{1 - \sum_b \gamma^{ab}\sigma^{ab}\sigma^{ba}n^b\chi^b}$ and variance $\frac{\sum_b(\sigma^{ab})^2n^bq^b}{\lrb{1 - \sum_b\Gamma^{ab}n^b\chi^b}^2}$, truncated to impose $x > 0$. With some straightforward substitutions from Eqs.~(13) we arrive at
\begin{align}
    \Prob(x|a) = \delta(x)\lrb{1 - \phi^a} + H(x)\frac{w_1(\Delta^a)}{M^a\sqrt{2\pi}} \explrbr{-\half \lrb{\frac{x w_1(\Delta^a)}{M^a} - \Delta^a}^2},
\end{align}
where $H(x)$ is the Heaviside step function. 

When we take the $B \to \infty$ limit and use the parameters of the cascade model, we arrive at Eq.~(29) of the main text
\begin{align}
    \Prob(x|\alpha) = \delta(x)\lrb{1 - \phi(\alpha)} + \frac{H(x)}{\sqrt{2\pi}}\frac{w_1[\Delta(\alpha)]}{M(\alpha)} \explrbr{-\half \lrb{\frac{x w_1[\Delta(\alpha)]}{M(\alpha)} - \Delta(\alpha)}^2}.
\end{align}
 To obtain a species abundance distribution, we find the probability that any species in the community has abundance $x$. We must therefore integrate the above over $\alpha$
\begin{align}
    \Prob(x) = \int_0^1d\alpha~n(\alpha)\Prob(x|\alpha).
\end{align}
The species abundance distribution is therefore
\begin{align}
    \Prob(x) = \delta(x)\lrb{1 - \E{\phi}} + \frac{H(x)}{\sqrt{2\pi}}\E{\frac{w_1(\Delta)}{M}\explrbr{-\half \lrb{\frac{x w_1(\Delta)}{M} - \Delta}^2}}, \label{app:eq:SAD}
\end{align}
where we recall the meaning of the average $\E{\dots}$ in \cref{app:eq:squarebracketaverage}. To compute the expression on the right-hand side of \cref{app:eq:SAD}, we first solve Eqs.~(26) for $\Delta_0, \Delta_1$ and $u$, then compute $M(\alpha)$ using either Eq.~(34) or \cref{app:eq:HAD}. Finally, we then use Eq.~(28) [equivalently \cref{app:eq:phi}] to obtain $\E{\phi}$. The integral
\begin{align}
    \E{\frac{w_1(\Delta)}{M}\explrbr{-\half \lrb{\frac{x w_1(\Delta)}{M} - \Delta}^2}} = \int_0^1 d\alpha~n(\alpha)\frac{w_1(\Delta(\alpha))}{M(\alpha)}\explrbr{-\half \lrb{\frac{x w_1(\Delta(\alpha))}{M(\alpha)} - \Delta(\alpha)}^2},
\end{align}
can be computed either by first finding $\Delta(\alpha)$ using Eq.~(23) [equivalently \cref{app:eq:DeltaODE}], or by changing variables to $\Delta$, again using Eq.~(23) [equivalently \cref{app:eq:DeltaODE}].
\subsubsection{\texorpdfstring{$P(x)$}{} in Eq.~(32) is a probability distribution}
Here we demonstrate that $P(x)$, as given in \cref{app:eq:SAD} [equivalently Eq.~(32)], is a probability distribution, that is, we show that $\int_{-\infty}^\infty dx~\Prob(x) = 1$. We start by demonstrating
\begin{align}
    \int_{-\infty}^\infty dx~\frac{H(x)}{\sqrt{2\pi}}\E{\frac{w_1(\Delta)}{M}\explrbr{-\half \lrb{\frac{x w_1(\Delta)}{M} - \Delta}^2}} = \E{\phi}.
\end{align}
This is done in the following steps
\begin{align}
        \E{\frac{1}{\sqrt{2\pi}}\int_0^\infty dx~\frac{w_1(\Delta)}{M}\explrbr{-\half \lrb{\frac{x w_1(\Delta)}{M} - \Delta}^2}}
        &= \E{\frac{1}{\sqrt{2\pi}}\int_{\Delta}^\infty dy~\explrbr{-\half y^2}} \nonumber \\
        &= \E{w_0(\Delta)} = \E{\phi},
\end{align}
where the substitution $y = x w_1(\Delta)/M - \Delta$ gives the first equality. The second equality is the definition of $w_0(\Delta)$. We therefore find that 
\begin{align}
    \int_{-\infty}^\infty dx~\Prob(x) = \lrb{1 - \E{\phi}} + \E{\phi} = 1,
\end{align}
as claimed.
\subsection{Rank abundance distributions (RADs)}\label{appendix:RAD}
As mentioned in the main text, a rank abundance distribution is obtained from ranking species by abundance, with the largest abundance species being given rank $0$ and lowest abundance species being given rank $1$. 

The function $\Prob(x)$ in \cref{app:eq:SAD} can be used to construct such a ranking: there are $\int_0^xdx'~\Prob(x')$ species with an abundance smaller than $x$. Hence a plot of abundance $x$ against $1 - \int_0^x dx'~\Prob(x')$ provides the rank abundance distribution.

\subsection{Derivation of Eq.~(34) and of survival distributions}
For a species with index $\alpha$ (recall that $\alpha = a/B$ in the $B \to \infty$ limit, see Section IV.2 in the main text), there are $N\int_0^\alpha d\beta~n(\beta)$ species lower in the hierarchy. Hence we use 
\begin{align}
    r(\alpha) \equiv \int_0^\alpha d\beta~n(\beta), \label{app:eq:r}
\end{align} 
as an alternative measure of a species position in the hierarchy. To find $M(\alpha)$, we use \cref{app:eq:FP3}, which we repeat here
\begin{align}
    M(\alpha) &= w_1[\Delta(\alpha)]\sigma \sqrt{q_\rho(\alpha)}.
\end{align}
Dividing the above by its average gives
\begin{align}
   \frac{M(\alpha)}{\E{M}} &= \frac{w_1[\Delta(\alpha)] \sqrt{q_\rho(\alpha)}}{\E{w_1(\Delta)\sqrt{q_\rho}}}.
\end{align}
Substituting for $q_\rho(\alpha)$ using \cref{app:eq:qrho} gives Eq.~(34) from the main text. Explicitly
\begin{align}
    \frac{M(\alpha)}{\E{M}} &= A w_1[\Delta(\alpha)]\explrbr{\frac{\ell \sigma^2\ln\rho}{u^2} \int_0^\alpha d\beta~n(\beta)w_2(\Delta(\beta))}, \label{app:eq:HAD}
\end{align}
where
\begin{align}
    A &= \int_0^1 d\alpha~n(\alpha)w_1(\Delta(\alpha))\explrbr{\frac{\ell \sigma^2\ln\rho}{u^2} \int_0^\alpha d\beta~n(\beta)w_2(\Delta(\beta))}.
\end{align}
To explicitly compute $A$, we first solve \cref{app:eq:FPeqnsCascade} for $\Delta_0, \Delta_1, u$, which we then use to find $\Delta(\alpha)$ with \cref{app:eq:DeltaODE} [the value of $u$ is substituted into $\theta(\Delta, u)$ and $\Delta_0, \Delta_1$ are the boundary conditions on $\Delta(\alpha)$]. Once $\Delta(\alpha)$ is known, we can explicitly compute the above integrals numerically. A parametric plot of $M(\alpha)$ against $r(\alpha)$ constitutes an HAD, as in panel (c) of Fig.~ 3 in the main text.

To produce a survival distribution we first find $\Delta(\alpha)$ for given parameters $\nu, \sigma, \rho, \gamma$ and then produce a parametric plot with $w_0[\Delta(\alpha)]$ on the $y$ axis and $r(\alpha)$ on the $x$ axis.

\section{Local Stability Analysis}
\label{appendix:FPStability}
\subsection{Derivation for block structured matrices}
We first find the local stability of the fixed point of the system with general block-structured interaction and then move on to the specific case of the cascade model. We follow along the lines of the stability analyses in \cite{1overf_noise,Galla_2018}.

The local stability of possible fixed points can be probed by addition of an infinitesimal independent and identically distributed Gaussian perturbation $\epsilon \xi^a(t)$ to each block in the effective dynamics Eq.~(6). In a stable regime we expect the system to return to the fixed point when perturbed. 

Applying these perturbations, we have 
\begin{align}
	\dot{x}^a(t) = x^a(t) \lrbracket{1 - x^a(t) + \sum_bn^b\lrsb{ \mu^{ab} M^b(t) +\gamma^{ab}\sigma^{ab}\sigma^{ba} \int_0^t dt' G^b(t,t') x^a(t')} + \eta^a(t) + \epsilon \xi^a(t)}.\label{app:eq:perturbed}
\end{align}
We quantify the linear perturbations of $x^a(t)$ and $\eta^a(t)$ about the fixed point (which we assume are of the order $\epsilon$) by $y^a(t), \kappa^a(t)$ respectively, such that
\begin{align}
	x^a(t) &= x_*^a + \epsilon y^a(t) \nonumber \\
	\eta^a(t) &= \eta_*^a + \epsilon \kappa^a(t).
\end{align}
We obtain the following self-consistency conditions [see \cref{app:eq:eta}]
\begin{align}
\avg{\kappa^a(t) \kappa^a(t')} &= \sum_b n^b(\sigma^{ab})^2 \avg{y^a(t) y^a(t')}.
\end{align}
Assuming time translation invariance in the long-time limit, linearising \cref{app:eq:perturbed} around the non-zero fixed point gives
\begin{align}
	\dot{y}^a(t) = x^a_* \lrbracket{- y^a(t) + \sum_bn^b \gamma^{ab}\sigma^{ab}\sigma^{ba}\int_0^t dt'~G^b(t - t') y^a(t')  + \kappa^a(t) + \xi^a(t)}. \label{app:eq:linAroundNonZeroFP}
\end{align}
  
We now follow \cite{1overf_noise} by going to Fourier space,
\begin{align}
    i\omega \hat{y}^a(\omega) = x_*^a\lrb{-\hat y^a(\omega) + \sum_bn^b\gamma^{ab}\sigma^{ab}\sigma^{ba}\hat G^b(\omega)\hat y^a(\omega) + \hat \kappa^a(\omega) + \hat \xi^a(\omega)}.
\end{align}
Squaring and averaging over $\kappa^a$ and $\xi^a$ we find
\begin{align}
    \lrb{\frac{|\omega|^2}{x_*^2} + \Big|1 - \sum_bn^b\gamma^{ab}\sigma^{ab}\sigma^{ba}\hat G^b(\omega)\Big|^2} \avg{|\hat y^a(\omega)|^2}  = \phi^a \lrb{\sum_bn^b \lrb{\sigma^{ab}}^2 \avg{|\hat y^b(\omega)|^2} + 1},
\end{align}
where the factor of $\phi^a$ is due to the fact that \cref{app:eq:linAroundNonZeroFP} only applies to non-zero fixed points, fluctuations around the zero point decay and hence do not contribute to $\avg{|\hat y^a(\omega)|^2}$. Noting that $\hat G^a(0) = \chi^a$, we now set $\omega = 0$ (see \cite{1overf_noise}) and find
\begin{align}
    \lrb{1 - \sum_bn^b\gamma^{ab}\sigma^{ab}\sigma^{ba}\chi^b}^2 Y^a &= \phi^a\lrb{\sum_bn^b \lrb{\sigma^{ab}}^2 Y^b + 1}, \label{app:eq:StabilityGeneralCond}
\end{align}
where $Y^a \equiv \avg{|\hat y^a(0)|^2}$. Assuming a stationary state in which $\avg{y(t)y(t + \tau)}$ depends on $\tau$ only, then $Y^a = \int d\tau~\avg{y(t)y(t + \tau)}$, and we may conclude that if $Y^a$ diverges, then perturbations do not decay to zero. Hence, a non-zero fixed point is unstable if the only solution to \cref{app:eq:StabilityGeneralCond} is one in which $Y^a$ diverges for some $a$. 
\subsection{Specific case of the cascade model}
We now obtain the stability criterion in the specific case of the Cascade model. We follow very similar lines to \cref{appendix:btoinfty} to find the $B \to \infty$ limit of \cref{app:eq:StabilityGeneralCond}. The result is
\begin{align}
    u(\alpha)^2 Y(\alpha) &= w_0[\Delta(\alpha)]\lrb{1 + \int_0^1d\beta~n(\beta)\sigma(\alpha, \beta)^2Y(\beta)},
\end{align}
which, again following very similar lines to \cref{appendix:btoinfty}, takes the following form in the cascade model
\begin{align}
	u^2 Y(\alpha) = w_0[\Delta(\alpha)] \lrb{\sigma^2\rho^2 \int_0^\alpha d\beta\ n(\beta)Y(\beta) + \frac{\sigma^2}{\rho^2} \int_\alpha^1 d\beta\ n(\beta)Y(\beta) + 1}.
\end{align}
This can be written as
\begin{align}
    u^2Y(\alpha) &= w_0[\Delta(\alpha)]\left(1+\sigma^2Y_\rho(\alpha)\right), \label{app:eq:StabilityCondOpperCascade} 
\end{align}
where we recall the definition in \cref{app:eq:f_mu_sigma}. From the general case considered in \cref{appendix:FPStability} we know that our system is linearly unstable if the solution $Y(\alpha)$ to the condition in \cref{app:eq:StabilityCondOpperCascade} is unbounded. For the purposes of our analysis we will assume the weaker condition that the integral $\E{Y}$ is unbounded, the justification for this assumption is the numerical agreement we find in \cref{appendix:numerical_verification_of_stability_curves}.

Following a very similar procedure to that in \cref{appendix:uRelationsDerivation}, we find
\begin{align}
    u^2 = \Lmbr{\frac{1}{\E{Y}} + \frac{\sigma^2}{\rho^2}, \frac{1}{\E{Y}} + \sigma^2\rho^2}\E{w_0(\Delta)}.
\end{align}
The following property of the logarithmic mean
\begin{align}
    \Lmbr{|a| + x, |a| + y} > \Lmbr{x, y},
\end{align}
tells us that, when $\E{Y}$ is finite, the following must hold
\begin{align}
    u^2 < \Lmbr{\frac{\sigma^2}{\rho^2}, \sigma^2\rho^2} \E{w_0} = \ell\sigma^2\E{w_0(\Delta)},
\end{align}
When $\E{Y}$ diverges, that is, on the edge of linear instability, the inequality becomes an equality, giving a sufficient condition for linear instability. When combined with \cref{app:eq:usig}, we see that the system is unstable to linear perturbation when
\begin{align}
    \E{w_0(\Delta)} \geq \E{w_2(\Delta)}. \label{app:eq:opper_for_w_funcs}
\end{align}
Plugging the above condition in to Eqs.~(26) [equivalently \cref{app:eq:FPeqnsCascade}] gives Eq.~(35) from the main text
\begin{align}\label{app:eq:opper}
    \sigma^2 \geq \frac{\ell}{(\ell + \gamma)^2}\frac{1}{\E{\phi}}.
\end{align}
Note that nothing in this derivation depended explicitly on $\mu(\alpha, \beta)$ (equivalently $\mu^{ab}$). Hence the choice of $\mu(\alpha, \beta)$ only enters into the stability analysis through its effect on average species survival rates $\E{\phi}$.
\subsection{Local Stability in the cases when \texorpdfstring{$\nu = 0$}{} or \texorpdfstring{$\rho = 1$}{} or both}\label{appendix:resilience_nu0}
In this section we will look at Eq.~(35) [equivalently \cref{app:eq:opper}] in three special cases. Firstly, when $\nu = 0$ and $\rho = 1$ the model reduces to that in \cite{Galla_2018}. In these references the equivalent of Eq.~(35) is derived [see Eq.~(13) in Ref. \cite{Galla_2018}]
\begin{align}
    \sigma^2_{\nu = 0,~\rho = 1} \geq \frac{2}{(1 + \gamma)^2}. \label{app:eq:oneblockopper}
\end{align}
By Eq.~(22) [equivalently \cref{app:eq:ell}], when $\rho = 1$ we have $\ell = 1$. Hence, by inspection of both \cref{app:eq:oneblockopper} and \cref{app:eq:opper}, we see that the average survival rate $\E{\phi}$ at the onset of linear instability in the case where $\nu = 0$ and $\rho = 1$ is equal to $1/2$.

We will now derive similar conditions in the cases when either $\nu \neq 0$ or $\rho \neq 1$. This will demonstrate that the critical value of $\sigma$ at which the system becomes unstable is larger with larger values of $|\nu|$, and smaller with smaller values of $|\rho - 1|$. That is, a larger value of $|\rho - 1|$ destabilises and larger $|\nu|$ stabilises.
\subsubsection{The case \texorpdfstring{$\nu = 0$}{}}
When $\nu = 0$ we find that Eq.~(24) [equivalently \cref{app:eq:thetadefn}] becomes
\begin{align}
    \theta(\Delta) = -\frac{u^2}{\ell \sigma^2 \ln\rho^2}\frac{1}{\Delta w_2(\Delta)},
\end{align}
so that Eqs.~(26) [equivalently \cref{app:eq:FPeqnsCascade}] are
\begin{align}
    \Delta_1 \rho^2 &= \Delta_0, \label{app:eq:FPEqnsrho1} \\
    \frac{1}{u} - 1 &= \frac{\gamma}{\ell \ln \rho^2}\int_{\Delta_1}^{\Delta_0}\frac{w_0(x)~dx}{xw_2(x)}, \label{app:eq:FPEqnsrho2} \\
    1 &= \frac{u^2}{\sigma^2 \ell \ln \rho^2}\int_{\Delta_1}^{\Delta_0} \frac{dx}{x w_2(x)}. \label{app:eq:FPEqnsrho3}
\end{align}
We want to solve these equations when the system is on the edge of linear instability, which, by \cref{app:eq:opper_for_w_funcs}, occurs when $\E{w_0(\Delta)} = \E{w_2(\Delta)}$. Using the identity $w_2(\Delta) = w_0(\Delta) + \Delta w_1(\Delta)$ [see \cref{app:eq:wdefn} for the definition of $w_0, w_1, w_2$  we may more conveniently write this condition as $\E{\Delta w_1(\Delta)} = 0$. We then conclude that the system is on the edge of instability when  \cref{app:eq:FPEqnsrho1,app:eq:FPEqnsrho2,app:eq:FPEqnsrho3} are satisfied, and when simultaneously  
\begin{align}
    \E{\Delta w_1(\Delta)} = \int_{\Delta_0}^{\Delta_1}dx~\theta(x)x w_1(x) = \frac{u^2}{\ell \sigma^2 \ln\rho^2}\int_{\Delta_1}^{\Delta_0}dx~\frac{w_1(x)}{w_2(x)} = 0.
\end{align}
As $w'_2(x) = 2 w_1(x)$, we can explicitly write the above as
\begin{align}
    \ln \frac{w_2(\Delta_0)}{w_2(\Delta_1)} = 0,
\end{align}
implying that $\Delta_0 = \Delta_1$. Combining this with \cref{app:eq:FPEqnsrho1} we see that both of $\Delta_0 = \Delta_1\rho^2$ and $\Delta_0 = \Delta_1$ must hold, and therefore $\Delta_0 = \Delta_1 = 0$ as we have assumed that $\rho \neq 1$.

For the remainder of this derivation we write $\Delta_1 = \Delta$ so that $\Delta_0 = \Delta \rho^2$. By the above considerations, we are interested in \cref{app:eq:FPEqnsrho2,app:eq:FPEqnsrho3} in the limit $\Delta \to 0$. Examining \cref{app:eq:FPEqnsrho2} we have
\begin{align}
    \frac{1}{u} - 1 = \frac{\gamma}{\ell \ln \rho^2}\lim_{\Delta \to 0}\int_{\Delta}^{\Delta\rho^2}\frac{w_0(x)~dx}{xw_2(x)} = \frac{\gamma}{\ell \ln \rho^2} \lim_{\Delta \to 0}\int_{\Delta}^{\Delta\rho^2} \frac{1}{x} + \mathcal{O}(1)~dx = \frac{\gamma}{\ell},
\end{align}
so that
\begin{align}
    u = \frac{1}{1 + \gamma/\ell}.
\end{align}
Similarly, from \cref{app:eq:FPEqnsrho3} we find
\begin{align}
    u^2 = \half \ell \sigma^2,
\end{align}
so that the equivalent of Eq.~(35) when $\nu = 0$ is
\begin{align}
    \sigma^2_{\nu = 0} \geq \frac{2\ell}{(\ell + \gamma)^2}. \label{app:eq:oppernu0}
\end{align}
In particular, we see that at the edge of instability, $\E\phi = 1/2$.

\cref{app:eq:oppernu0} tells us that, when $\nu = 0$, increasing $|\rho - 1|$ is always destabilising. To see this, we look at the sign of the derivative of $\sigma^2$ with respect to $\rho$. Taking the logarithmic derivative of both sides of \cref{app:eq:oppernu0} gives
\begin{align}
    \frac{1}{\sigma^2}\frac{(\sigma^2)'}{\ell'} &= \frac{1}{\ell} - 2\frac{1}{\ell + \gamma} \leq 0.
\end{align}
The last inequality follows from
\begin{align}
    \frac{1}{1 + \gamma/\ell} \geq \frac 12,
\end{align}
which itself follows from the fact that $\gamma \in [-1, 1]$ and $\ell \in [1, \infty)$, so that $\gamma/\ell \in [-1, 1]$. Therefore, we find
\begin{align}
    \sign[(\sigma^2)'] = -\sign(\ell'),
\end{align}
it is easily verified that $\ell$ increases if $|\rho - 1|$ is increased, and so therefore the critical value of $\sigma^2$ decreases if $|\rho - 1|$ is increased. Hence, in the case $\nu = 0$ increasing $|\rho - 1|$ always moves the system towards linear instability.
\subsubsection{The case \texorpdfstring{$\rho = 1$}{}}
When $\rho = 1$, Eq.~(24) becomes
\begin{align}
    \theta(\Delta) = \frac{u}{2\nu}\frac{1}{w_1(\Delta)},
\end{align}
so that Eqs.~(26) are
\begin{align}
    1 &= \frac{\sigma^2}{2\nu u}\lrb{\ln\lrb{\frac{w_1(\Delta_1)}{w_1(\Delta_0)}} + \half \lrb{\Delta_1^2 - \Delta_0^2}}, \label{app:eq:FPEqnsnu1}\\
    1 - u &= \frac{\gamma \sigma^2}{2\nu}\ln\lrb{\frac{w_1(\Delta_1)}{w_1(\Delta_0)}}, \label{app:eq:FPEqnsnu2}\\
    1 &= \frac{u}{2 \nu} \int_{\Delta_0}^{\Delta_1} \frac{dx}{w_1(x)}. \label{app:eq:FPEqnsnu3}
\end{align}
As with the case of $\nu = 0$, we know that the system is on the edge of linear instability when $\E{\Delta w_1(\Delta)} = 0$, therefore
\begin{align}
    \int_{\Delta_0}^{\Delta_1} \frac{x w_1(x)\ dx}{w_1(x)} = 0,
\end{align}
as this integral simply evaluates to $1/2(\Delta_1^2 - \Delta_0^2)$, we conclude that $\Delta_0 = \pm \Delta_1$. Suppose $\Delta_0 = \Delta_1$, then by \cref{app:eq:FPEqnsnu3}, $u$ must diverge, which is inconsistent with \cref{app:eq:FPEqnsnu2}. Hence, the only consistent choice is $\Delta_0 = -\Delta_1$, and so we call $\Delta_1 = \Delta$ and $\Delta_0 = -\Delta$. Eliminating $u$ from \cref{app:eq:FPEqnsnu1,app:eq:FPEqnsnu2,app:eq:FPEqnsnu3} then gives
\begin{align}
    \sigma^2_{\rho = 1} &= \frac{1}{(1 + \gamma)^2}\left.\int_{-\Delta}^\Delta \frac{dx}{w_1(x)}\right/\int_{-\Delta}^{\Delta}dx~\frac{w_0(x)}{w_1(x)}, \label{app:eq:opperrho11} \\
    \nu &= \half \frac{1}{1 + \gamma} \int_{-\Delta}^\Delta \frac{dx}{w_1(x)}. \label{app:eq:opperrho12}
\end{align}
The above are unchanged under the transformation $\nu \to -\nu,~\Delta \to -\Delta$, so for the remainder of this section we will take $\nu > 0,~\Delta > 0$ without loss of generality, one can make the same conclusions for $\nu < 0$ by reversing the sign of $\Delta$. Taking $\Delta > 0$ allows us to use some useful properties of the functions $w_0(\Delta)$ and $w_1(\Delta)$ in the following.

We will look at the logarithmic derivative of $\sigma^2$ with respect to $\nu$ and show that this quantity is positive. Firstly, we note that $w_1(x) > 0$ for all $x$, implying that $\nu$ is an increasing function of $\Delta$ by \cref{app:eq:opperrho12}. In turn this implies that the derivative of $\sigma^2$ with respect to $\nu$ has the same sign as the derivative of $\sigma^2$ with respect to $\Delta$ by the chain rule. Hence, it is enough to show that the following is positive
\begin{align}
    \diff[\ln\sigma^2]{\Delta} &= \frac{1/w_1(\Delta) + 1/w_1(-\Delta)}{\int_{-\Delta}^{\Delta}\frac{dx}{w_1(x)}} - \frac{w_0(\Delta)/w_1(\Delta) + w_0(-\Delta)/w_1(-\Delta)}{\int_{-\Delta}^{\Delta}\frac{dx~w_0(x)}{w_1(x)}}. \label{app:eq:opperClaimNu}
\end{align}
First, we show that for $\Delta \geq 0$
\begin{align}
    \int_{-\Delta}^\Delta \frac{dx~w_0(x)}{w_1(x)} \leq \half \int_{-\Delta}^\Delta \frac{dx}{w_1(x)}, \label{app:eq:w0Inequality}
\end{align}
which follows from the following observations. Firstly, $w_0(\Delta) - 1/2$ is an odd increasing function of $\Delta$ and is positive for $\Delta \geq 0$. Secondly, $1/w_1(\Delta)$ is a decreasing function of $\Delta$. Hence we may write
\begin{align}
    \int_0^\Delta dx~\frac{w_0(x) - 1/2}{w_1(x)} \leq \int_0^\Delta dx~\frac{w_0(x) - 1/2}{w_1(-x)},
\end{align}
or, using that $w_0(\Delta) - 1/2$ is odd
\begin{align}
    2\int_{-\Delta}^\Delta dx~\frac{w_0(x) - 1/2}{w_1(x)} \leq 0,
\end{align}
from which our claim follows.

We can re-write the claim in \cref{app:eq:opperClaimNu} as
\begin{align}
    \lrb{\frac{1}{w_1(\Delta)} + \frac{1}{w_1(-\Delta)}}\int_{-\Delta}^{\Delta}\frac{dx~w_0(x)}{w_1(x)} - \lrb{\frac{w_0(\Delta)}{w_1(\Delta)} + \frac{w_0(-\Delta)}{w_1(-\Delta)}}\int_{-\Delta}^{\Delta}\frac{dx}{w_1(x)} \geq 0.     
\end{align}
We now use \cref{app:eq:w0Inequality} to write
\begin{multline}
    \lrb{\frac{1}{w_1(\Delta)} + \frac{1}{w_1(-\Delta)}}\int_{-\Delta}^{\Delta}\frac{dx~w_0(x)}{w_1(x)} - \lrb{\frac{w_0(\Delta)}{w_1(\Delta)} + \frac{w_0(-\Delta)}{w_1(-\Delta)}}\int_{-\Delta}^{\Delta}\frac{dx}{w_1(x)} \\ \geq
    \lrcb{\lrb{\frac{1}{w_1(\Delta)} + \frac{1}{w_1(-\Delta)}} - 2\lrb{\frac{w_0(\Delta)}{w_1(\Delta)} + \frac{w_0(-\Delta)}{w_1(-\Delta)}}}\int_{-\Delta}^{\Delta}\frac{dx}{w_1(x)}.
\end{multline}
We now show that the quantity in the curly brackets is positive with the following manipulations
\begin{align}
    \lrb{\frac{1}{w_1(\Delta)} + \frac{1}{w_1(-\Delta)}} - 2\lrb{\frac{w_0(\Delta)}{w_1(\Delta)} + \frac{w_0(-\Delta)}{w_1(-\Delta)}} &= 
    \frac{1 - 2w_0(\Delta)}{w_1(\Delta)} - \frac{1 - 2w_0(-\Delta)}{w_1(-\Delta)} \nonumber \\
    &= \lrb{1 - 2w_0(\Delta)}\lrb{\frac{1}{w_1(-\Delta)} - \frac{1}{w_1(\Delta)}},
\end{align}
which is positive for positive $\Delta$, as $w_0(\Delta) \geq 1/2$ and $1/w_1(\Delta)$ is a decreasing function. Hence, $\sigma^2_{\rho = 1}$ is an increasing function of $\nu$, in other words, increasing $\nu$ stabilises the system. 
\section{Verification of the criteria for stability using computer simulation}
\label{appendix:numerical_verification_of_stability_curves}
\begin{figure*}[!ht]
    \centering
    \makebox[\textwidth][c]{\includegraphics[width=\textwidth]{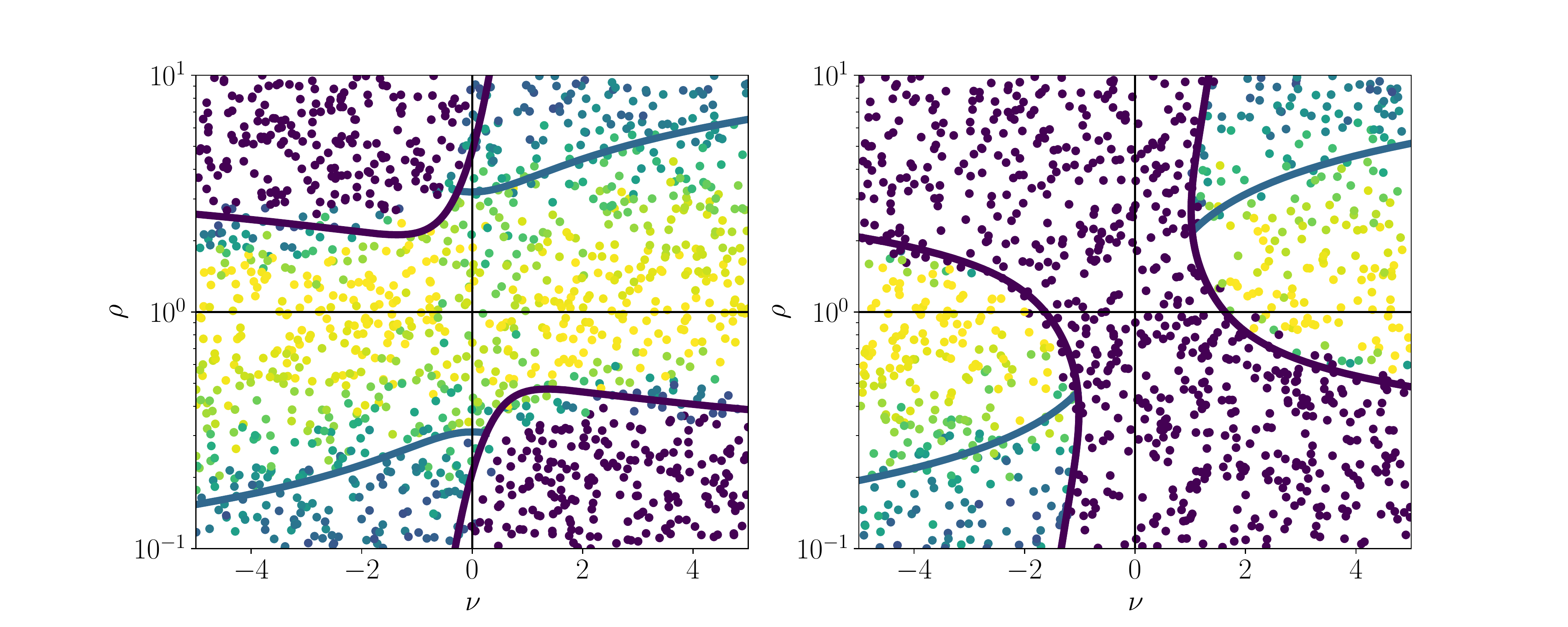}}
    \caption{Simulation data for central and right plots of Fig. 6 in the main text. Parameters are $\mu = -0.5, \sigma = 0.7, \gamma = -0.8$ for the left plot and $\mu = 0.5, \sigma = 0.8, \gamma = -0.8$ for the right plot. Each data point is stability data for a community of $N = 200$ species, averaged over $20$ runs, $(\nu, \rho)$ pairs are randomly sampled. Points are colored according to their measured stability, purple (darkest) indicates that abundances diverge, blue (medium darkness) indicates linear instability and yellow (lightest) indicates stability. There is a colour gradient between stable and linearly unstable as the instability boundary is only perfectly sharp in the $N \to \infty$ limit.}
    \label{app:fig:nu-rho-stability-proof}
\end{figure*}
In order to test the stability criteria in Eqs.~(35) and (36), we randomly sample a large number of pairs $(\nu, \rho)$. $\nu$ is sampled uniformly from the range $(-5, 5)$ and $\rho$ is obtained as $\rho=\exp(X)$, where $X$ is uniformly sampled in the range $(-\ln 10, \ln 10)$. This ensures that the points shown in \cref{app:fig:nu-rho-stability-proof} are uniformly distributed on the logarithmic scale. The values of $\mu, \sigma, \gamma$ are fixed. For each pair $(\nu, \rho)$ we simulate the system $20$ times with input parameters $\mu, \nu, \sigma, \rho, \gamma$, and determine an approximate probability of the systems stability, or lack thereof. Each point is then plotted in the $\nu - \rho$ plane, and given a colour representative of this probability, for example, if 10 of the twenty runs are deemed stable and 10 linearly unstable, then that particular point will have a colour half way between the stable and unstable colours on the colour map we have chosen. The result, as well as comparison to theory lines predicted by Eqs.~(35) and (36), are shown in two particular instances in \cref{app:fig:nu-rho-stability-proof}.
\section{Independence of results from \texorpdfstring{$n(\alpha)$}{}}
\label{appendix:nalphaIndependence}
Here we prove the following statement about our system: Let $f(\alpha)$ be some function which does not explicitly depend on the relative number of species with index $\alpha$ [$n(\alpha)$]. We also imagine that $f(\alpha) = g[\Delta(\alpha)]$ for some function $g$. When both of these conditions are satisfied, then $\E{f}$ (equivalently $\E{g(\Delta)}$) is independent of $n(\alpha)$. 

To see this, we use Eq.~(23) [equivalently \cref{app:eq:DeltaODE}] to change variables $\alpha \to \Delta$ in the following
\begin{align}
    \E{f} = \int_0^1d\alpha~n(\alpha)f(\alpha) = \int_{\Delta_0}^{\Delta_1}d\Delta~\theta(\Delta, u)g(\Delta).
\end{align}
We recall Eqs.~(26) [equivalently \cref{app:eq:FPeqnsCascade}], which allow us to compute $\Delta_0, \Delta_1$ and $u$ as functions of $\nu, \sigma, \rho, \gamma$ only. Therefore, the integral over delta in the above depends only on $\nu, \sigma, \rho, \gamma$, as well as any parameters which $g$ depends on, which we have assumed excludes $n(\alpha)$. 
\subsection{\texorpdfstring{$\E{M}$}{} and \texorpdfstring{$\E{\phi}$}{} [Eqs.~(27) and (28)]}
In light of the preceding subsection, $\E{M}$ and $\E{\phi}$ must be independent of $n(\alpha)$ by Eqs.~(27) and (28) [equivalently \cref{app:eq:M,app:eq:phi}]. Here, we give an alternative proof by demonstrating that $M(\alpha) = M[\Delta(\alpha)]$ and $\phi(\alpha) = \phi[\Delta(\alpha)]$. We recall \cref{app:eq:HAD}, 
\begin{align}
    \frac{M(\alpha)}{\E{M}} &= A w_1(\Delta(\alpha))\explrbr{\frac{\ell \sigma^2\ln\rho}{u^2} \int_0^\alpha d\beta~n(\beta)w_2(\Delta(\beta))}, 
\end{align}
changing variables with \cref{app:eq:DeltaODE} in the exponentiated integral gives
\begin{align}
    \int_0^\alpha d\beta~n(\beta)w_2(\Delta(\beta)) = \int_{\Delta_0}^{\Delta(\alpha)} d\Delta~\theta(\Delta, u)w_2(\Delta),
\end{align}
demonstrating that $M(\alpha) = M(\Delta(\alpha))$, so that $\E{M}$ is independent of $n(\alpha)$. Noting that $\phi(\alpha) = w_0(\Delta(\alpha))$ gives the corresponding claim for $\E{\phi}$.
\subsection{SADs [Eq.~(32)]}
This follows from changing variables $\alpha \to \Delta$ in the integral in the second term in \cref{app:eq:SAD}. We find
\begin{align}
    \E{\frac{w_1(\Delta)}{M}\explrbr{-\half \lrb{\frac{x w_1(\Delta)}{M} - \Delta}^2}} = 
    \int_0^1 d\Delta~\theta(\Delta, u)\frac{w_1(\Delta)}{M(\Delta)}\explrbr{-\half \lrb{\frac{x w_1(\Delta)}{M(\Delta)} - \Delta}^2},
\end{align}
where, by \cref{app:eq:HAD,app:eq:DeltaODE}, we have
\begin{align}
    \frac{M(\Delta)}{\E{M}} = \frac{w_1(\Delta)\explrbr{\frac{\ell \sigma^2\ln\rho}{u^2} \int_{\Delta_0}^\Delta d\Delta'~\theta(\Delta', u)w_2(\Delta')}}{\int_{\Delta_0}^{\Delta_1}d\Delta'~w_1(\Delta')\explrbr{\frac{\ell \sigma^2\ln\rho}{u^2} \int_{\Delta_0}^{\Delta'} d\Delta''~\theta(\Delta'', u)w_2(\Delta'')}},
\end{align}
which has no $n(\alpha)$ dependence, hence $\Prob(x)$ in Eq.~(32) [equivalently \cref{app:eq:SAD}] also has no dependence on $n(\alpha)$.
\subsection{RADs}
This follows from the independence of $\Prob(x)$ from $n(\alpha)$, as this function is used to define an RAD (see \cref{appendix:RAD}.
\subsection{HADs [Eq.~(34)]}
\label{appendix:HADsAreNalphaIndependent}
To demonstrate that HADs and survival distributions are independent of $n(\alpha)$, we will derive explicit expressions for $M$ and $\phi$ as functions of $r$, where $r$ is given by \cref{app:eq:r}. First we note that \cref{app:eq:DeltaODE} can be written as
\begin{align}
    \diff[\Delta]{r} &= \frac{1}{\theta(\Delta, u)},
\end{align}
and therefore $\Delta(r)$ is independent of $n(\alpha)$. Further, we can now write
\begin{align}
    \frac{M(r)}{\E{M}} &= A w_1[\Delta(r)] \explrbr{\frac{\ell\sigma^2\ln\rho}{u^2} \int_0^r dr'~w_2[\Delta(r')]},
\end{align}
where 
\begin{align}
    A &= \int_0^1 dr~w_1[\Delta(r)]\explrbr{\frac{\ell \sigma^2\ln\rho}{u^2} \int_0^r dr'~w_2[\Delta(r')]},
\end{align}
and hence $M(r)$ has no $n(\alpha)$ dependence.
\subsection{Survival Distributions}
Similarly to $M(r)$, we consider $\phi(r) = w_0[\Delta(r)]$, which also has no $n(\alpha)$ dependence, as $\Delta(r)$ has none.
\subsection{Linear Instability [Eq.~(35)]}
We recall Eq.~(35) [equivalently \cref{app:eq:opper}]
\begin{align}
    \sigma^2 \geq \frac{\ell}{(\ell + \gamma)^2}\frac{1}{\E{\phi}},
\end{align}
which is independent of $n(\alpha)$ as $\ell$ depends only on $\rho$ [Eq.~(22)] and $\E{\phi}$ is independent of $n(\alpha)$.
\subsection{Diverging abundances instability [Eq.~(36)]}
This follows from the independence of $\E{M}$ on $n(\alpha)$, hence the point at which $\E{M} \to \infty$ is also independent of the choice of the function $n(\alpha)$. Alternatively, recall Eq.~(36) from the main text, abundances will diverge provided
\begin{align}
    \mu \geq \frac{\Delta_1\rho^2 + \Delta_0}{\Delta_1\rho^2 - \Delta_0}\nu,
\end{align}
and, by Eqs.~(26) [alternatively \cref{app:eq:FPeqnsCascade}], both of $\Delta_0$ and $\Delta_1$ are independent of $n(\alpha)$.

\bibliographystyle{unsrtnat}

\begin{thebibliography}{57}
\providecommand{\natexlab}[1]{#1}
\providecommand{\url}[1]{\texttt{#1}}
\expandafter\ifx\csname urlstyle\endcsname\relax
  \providecommand{\doi}[1]{doi: #1}\else
  \providecommand{\doi}{doi: \begingroup \urlstyle{rm}\Url}\fi

\bibitem[Gardner and Asgby(1970)]{GARDNER1970}
M.~R. Gardner and W.~R. Asgby.
\newblock Connectance of large dynamic (cybernetic) systems: Critical values
  for stability.
\newblock \emph{Nature}, 228\penalty0 (5273):\penalty0 784--784, Nov 1970.
\newblock ISSN 1476-4687.
\newblock \doi{10.1038/228784a0}.

\bibitem[May(1972)]{MAY1972}
R.~M. May.
\newblock Will a large complex system be stable?
\newblock \emph{Nature}, 238\penalty0 (5364):\penalty0 413--414, Aug 1972.
\newblock ISSN 1476-4687.
\newblock \doi{10.1038/238413a0}.

\bibitem[Dunne et~al.(2002)Dunne, Williams, and Martinez]{Dunne12917}
J.~A. Dunne, R.~J. Williams, and N.~D. Martinez.
\newblock Food-web structure and network theory: The role of connectance and
  size.
\newblock \emph{Proceedings of the National Academy of Sciences}, 99\penalty0
  (20):\penalty0 12917--12922, 2002.
\newblock ISSN 0027-8424.
\newblock \doi{10.1073/pnas.192407699}.

\bibitem[Pimm et~al.(1991)Pimm, Lawton, and Cohen]{Food_Web_Structure}
S.~L. Pimm, J.~H. Lawton, and J.~E. Cohen.
\newblock Food web patterns and their consequences.
\newblock \emph{Nature}, 350\penalty0 (6320):\penalty0 669--674, Apr 1991.
\newblock ISSN 1476-4687.
\newblock \doi{10.1038/350669a0}.

\bibitem[MacArthur(1955)]{MacArthur_1955}
R.~MacArthur.
\newblock Fluctuations of animal populations and a measure of community
  stability.
\newblock \emph{Ecology}, 36\penalty0 (3):\penalty0 533--536, 1955.
\newblock \doi{https://doi.org/10.2307/1929601}.

\bibitem[McCann(2000)]{McCann2000}
K.~S. McCann.
\newblock The diversity--stability debate.
\newblock \emph{Nature}, 405\penalty0 (6783):\penalty0 228--233, May 2000.
\newblock ISSN 1476-4687.
\newblock \doi{10.1038/35012234}.

\bibitem[Grimm and Wissel(1997)]{Grimm1997}
V.~Grimm and C.~Wissel.
\newblock Babel, or the ecological stability discussions: an inventory and
  analysis of terminology and a guide for avoiding confusion.
\newblock \emph{Oecologia}, 109\penalty0 (3):\penalty0 323--334, Feb 1997.
\newblock ISSN 1432-1939.
\newblock \doi{10.1007/s004420050090}.

\bibitem[Allesina and Tang(2015)]{Stability_Complexity_review_allesina}
S.~Allesina and S.~Tang.
\newblock The stability–complexity relationship at age 40: a random matrix
  perspective.
\newblock \emph{Population Ecology}, 57\penalty0 (1):\penalty0 63--75, 2015.
\newblock \doi{https://doi.org/10.1007/s10144-014-0471-0}.

\bibitem[Landi et~al.(2018)Landi, Minoarivelo, Br\"annstr\"om, Hui, and
  Dieckmann]{Landi_review}
P.~Landi, H.~Minoarivelo, Å. Br\"annstr\"om, C.~Hui, and U.~Dieckmann.
\newblock Complexity and stability of ecological networks: a review of the
  theory.
\newblock \emph{Population Ecology}, 07 2018.
\newblock \doi{10.1007/s10144-018-0628-3}.

\bibitem[Jacquet et~al.(2016)Jacquet, Moritz, Morissette, Legagneux, Massol,
  Archambault, and Gravel]{Jacquet2016}
C.~Jacquet, C.~Moritz, L.~Morissette, P.~Legagneux, F.~Massol, P.~Archambault,
  and D.~Gravel.
\newblock No complexity--stability relationship in empirical ecosystems.
\newblock \emph{Nature Communications}, 7\penalty0 (1):\penalty0 12573, Aug
  2016.
\newblock ISSN 2041-1723.
\newblock \doi{10.1038/ncomms12573}.

\bibitem[Girko(1985)]{girko1985circular}
V.~L. Girko.
\newblock Circular law.
\newblock \emph{Theory of Probability \& Its Applications}, 29\penalty0
  (4):\penalty0 694--706, 1985.
\newblock \doi{10.1137/1129095}.

\bibitem[Grilli et~al.(2016)Grilli, Rogers, and Allesina]{Grilli2016}
J.~Grilli, T.~Rogers, and S.~Allesina.
\newblock Modularity and stability in ecological communities.
\newblock \emph{Nature Communications}, 7\penalty0 (1):\penalty0 12031, Jun
  2016.
\newblock ISSN 2041-1723.
\newblock \doi{10.1038/ncomms12031}.

\bibitem[Allesina and Tang(2012)]{Allesina2012}
S.~Allesina and S.~Tang.
\newblock Stability criteria for complex ecosystems.
\newblock \emph{Nature}, 483\penalty0 (7388):\penalty0 205--208, Mar 2012.
\newblock ISSN 1476-4687.
\newblock \doi{10.1038/nature10832}.

\bibitem[Baron and Galla(2020)]{Baron2020Diffusion}
J.~W. Baron and T.~Galla.
\newblock Dispersal-induced instability in complex ecosystems.
\newblock \emph{Nature Communications}, 11\penalty0 (1):\penalty0 6032, Nov
  2020.
\newblock ISSN 2041-1723.
\newblock \doi{10.1038/s41467-020-19824-4}.

\bibitem[Gravel et~al.(2016)Gravel, Massol, and Leibold]{gravel2016stability}
D.~Gravel, F.~Massol, and M.~Leibold.
\newblock Stability and complexity in model meta-ecosystems.
\newblock \emph{Nature Communications}, 7:\penalty0 12457, 08 2016.
\newblock \doi{10.1038/ncomms12457}.

\bibitem[Gross et~al.(2009)Gross, Rudolf, Levin, and
  Dieckmann]{gross2009generalized}
T.~Gross, L.~Rudolf, S.~A. Levin, and U.~Dieckmann.
\newblock Generalized models reveal stabilizing factors in food webs.
\newblock \emph{Science}, 325\penalty0 (5941):\penalty0 747--750, 2009.
\newblock \doi{10.1126/science.1173536}.

\bibitem[Berlow et~al.(2004)Berlow, Neutel, Cohen, De~Ruiter, Ebenman,
  Emmerson, Fox, Jansen, Iwan~Jones, Kokkoris, Logofet, McKane, Montoya, and
  Petchey]{berlow2004interaction}
E.~L. Berlow, A.~M. Neutel, J.~E. Cohen, P.~C. De~Ruiter, B.~Ebenman,
  M.~Emmerson, J.~W. Fox, V.~A.~A. Jansen, J.~Iwan~Jones, G.~D. Kokkoris, D.~O.
  Logofet, A.~J. McKane, J.~M. Montoya, and O.~Petchey.
\newblock Interaction strengths in food webs: issues and opportunities.
\newblock \emph{Journal of Animal Ecology}, 73\penalty0 (3):\penalty0 585--598,
  2004.
\newblock \doi{https://doi.org/10.1111/j.0021-8790.2004.00833.x}.

\bibitem[Barab{\'a}s et~al.(2017)Barab{\'a}s, Michalska-Smith, and
  Allesina]{barabas2017self}
G.~Barab{\'a}s, M.~J. Michalska-Smith, and S.~Allesina.
\newblock Self-regulation and the stability of large ecological networks.
\newblock \emph{Nature Ecology {\&} Evolution}, 1\penalty0 (12):\penalty0
  1870--1875, Dec 2017.
\newblock ISSN 2397-334X.
\newblock \doi{10.1038/s41559-017-0357-6}.

\bibitem[Stone(2018)]{stone2018feasibility}
L.~Stone.
\newblock The feasibility and stability of large complex biological networks: a
  random matrix approach.
\newblock \emph{Scientific Reports}, 8\penalty0 (1):\penalty0 8246, May 2018.
\newblock ISSN 2045-2322.
\newblock \doi{10.1038/s41598-018-26486-2}.

\bibitem[Gibbs et~al.(2018)Gibbs, Grilli, Rogers, and
  Allesina]{gibbs2018effect}
Theo Gibbs, Jacopo Grilli, Tim Rogers, and Stefano Allesina.
\newblock Effect of population abundances on the stability of large random
  ecosystems.
\newblock \emph{Physical Review E}, 98\penalty0 (2), aug 2018.
\newblock \doi{10.1103/physreve.98.022410}.

\bibitem[Galla(2018)]{Galla_2018}
T.~Galla.
\newblock Dynamically evolved community size and stability of random
  lotka-volterra ecosystems.
\newblock \emph{{EPL} (Europhysics Letters)}, 123\penalty0 (4):\penalty0 48004,
  sep 2018.
\newblock \doi{10.1209/0295-5075/123/48004}.

\bibitem[Baron et~al.(2022{\natexlab{a}})Baron, Jewell, Ryder, and
  Galla]{baron2022non}
Joseph~W. Baron, Thomas~Jun Jewell, Christopher Ryder, and Tobias Galla.
\newblock Non-gaussian random matrices determine the stability of
  lotka-volterra communities, 2022{\natexlab{a}}.
\newblock URL \url{https://arxiv.org/abs/2202.09140}.

\bibitem[Bunin(2017)]{BuninOpenLVDynamics}
G.~Bunin.
\newblock Ecological communities with lotka-volterra dynamics.
\newblock \emph{Phys. Rev. E}, 95:\penalty0 042414, Apr 2017.
\newblock \doi{10.1103/PhysRevE.95.042414}.

\bibitem[Biroli et~al.(2018)Biroli, Bunin, and Cammarota]{biroli2018marginally}
G.~Biroli, G.~Bunin, and C.~Cammarota.
\newblock Marginally stable equilibria in critical ecosystems.
\newblock \emph{New Journal of Physics}, 20\penalty0 (8):\penalty0 083051, aug
  2018.
\newblock \doi{10.1088/1367-2630/aada58}.

\bibitem[Altieri et~al.(2021)Altieri, Roy, Cammarota, and
  Biroli]{altieri2021properties}
Ada Altieri, Felix Roy, Chiara Cammarota, and Giulio Biroli.
\newblock Properties of equilibria and glassy phases of the random
  lotka-volterra model with demographic noise.
\newblock \emph{Physical Review Letters}, 126\penalty0 (25), jun 2021.
\newblock \doi{10.1103/physrevlett.126.258301}.

\bibitem[Allesina et~al.(2015)Allesina, Grilli, Barab{\'a}s, Tang, Aljadeff,
  and Maritan]{Barabas_Cascade}
S.~Allesina, J.~Grilli, G.~Barab{\'a}s, S.~Tang, J.~Aljadeff, and A.~Maritan.
\newblock Predicting the stability of large structured food webs.
\newblock \emph{Nature Communications}, 6\penalty0 (1):\penalty0 7842, Jul
  2015.
\newblock ISSN 2041-1723.
\newblock \doi{10.1038/ncomms8842}.

\bibitem[Cohen et~al.(1990)Cohen, Luczak, Newman, and Zhou]{LVCM}
J.~E. Cohen, T.~Luczak, C.~M. Newman, and Z.~M Zhou.
\newblock Stochastic structure and nonlinear dynamics of food webs: qualitative
  stability in a lotka-volterra cascade model.
\newblock \emph{Proceedings of the Royal Society B}, 240\penalty0
  (1299):\penalty0 607--627, June 1990.
\newblock \doi{10.1098/rspb.1990.0055}.

\bibitem[De~Dominicis(1978)]{De_Dominics1978}
C.~De~Dominicis.
\newblock Dynamics as a substitute for replicas in systems with quenched random
  impurities.
\newblock \emph{Phys. Rev. B}, 18, 11 1978.
\newblock \doi{10.1103/PhysRevB.18.4913}.

\bibitem[Tang et~al.(2014)Tang, Pawar, and Allesina]{Tang_correlation}
S.~Tang, S.~Pawar, and S.~Allesina.
\newblock Correlation between interaction strengths drives stability in large
  ecological networks.
\newblock \emph{Ecology Letters}, 17\penalty0 (9):\penalty0 1094--1100, 2014.
\newblock \doi{https://doi.org/10.1111/ele.12312}.

\bibitem[Barbier et~al.(2018)Barbier, Arnoldi, Bunin, and Loreau]{Barbier2156}
M.~Barbier, J.F. Arnoldi, G.~Bunin, and M.~Loreau.
\newblock Generic assembly patterns in complex ecological communities.
\newblock \emph{Proceedings of the National Academy of Sciences}, 115\penalty0
  (9):\penalty0 2156--2161, 2018.
\newblock ISSN 0027-8424.
\newblock \doi{10.1073/pnas.1710352115}.

\bibitem[Barbier and Arnoldi(2017)]{Barbier_cavity_method}
M.~Barbier and J.~Arnoldi.
\newblock The cavity method for community ecology.
\newblock \emph{bioRxiv}, 2017.
\newblock \doi{10.1101/147728}.

\bibitem[Sidhom and Galla(2020)]{Laura_2020}
L.~Sidhom and T.~Galla.
\newblock Ecological communities from random generalized lotka-volterra
  dynamics with nonlinear feedback.
\newblock \emph{Physical Review E}, 101\penalty0 (3), Mar 2020.
\newblock ISSN 2470-0053.
\newblock \doi{10.1103/physreve.101.032101}.

\bibitem[Allesina(2020)]{allesina_GLVTour}
S.~Allesina.
\newblock A tour of the generalized lotka-volterra model, 2020.
\newblock URL \url{https://stefanoallesina.github.io/Sao_Paulo_School/}.

\bibitem[Kondoh(2003)]{Kondoh_GLV}
M.~Kondoh.
\newblock Foraging adaptation and the relationship between food-web complexity
  and stability.
\newblock \emph{Science}, 299\penalty0 (5611):\penalty0 1388--1391, 2003.
\newblock \doi{10.1126/science.1079154}.

\bibitem[M{\'e}zard et~al.(1987)M{\'e}zard, Parisi, and Virasoro]{mezard1987}
M.~M{\'e}zard, G.~Parisi, and M.~Virasoro.
\newblock \emph{Spin glass theory and beyond: An Introduction to the Replica
  Method and Its Applications}, volume~9.
\newblock World Scientific Publishing Company, London, 1987.

\bibitem[Coolen(2001)]{BiolHandbookGF}
A.~C.~C. Coolen.
\newblock \emph{Handbook of Biological Physics}, volume~4.
\newblock Elsevier Science B.V., 2001.

\bibitem[Opper and Diederich(1992)]{1overf_noise}
M.~Opper and S.~Diederich.
\newblock Phase transition and 1/f noise in a game dynamical model.
\newblock \emph{Phys. Rev. Lett.}, 69:\penalty0 1616--1619, Sep 1992.
\newblock \doi{10.1103/PhysRevLett.69.1616}.

\bibitem[Bahri et~al.(2020)Bahri, Kadmon, Pennington, Schoenholz,
  Sohl-Dickstein, and Ganguli]{Ganguli_deep_learning_review}
Y.~Bahri, J.~Kadmon, J.~Pennington, S.~S. Schoenholz, J.~Sohl-Dickstein, and
  S.~Ganguli.
\newblock Statistical mechanics of deep learning.
\newblock \emph{Annual Review of Condensed Matter Physics}, 11\penalty0
  (1):\penalty0 501--528, 2020.
\newblock \doi{10.1146/annurev-conmatphys-031119-050745}.

\bibitem[Galla and Farmer(2013)]{Galla1232}
T.~Galla and J.~D. Farmer.
\newblock Complex dynamics in learning complicated games.
\newblock \emph{Proceedings of the National Academy of Sciences}, 110\penalty0
  (4):\penalty0 1232--1236, 2013.
\newblock ISSN 0027-8424.
\newblock \doi{10.1073/pnas.1109672110}.

\bibitem[Baron et~al.(2022{\natexlab{b}})Baron, Jewell, Ryder, and
  Galla]{baron2021eigenvalues}
J.~W. Baron, T.~J. Jewell, C.~Ryder, and T.~Galla.
\newblock Eigenvalues of random matrices with generalized correlations: A path
  integral approach.
\newblock \emph{Phys. Rev. Lett.}, 128:\penalty0 120601, Mar
  2022{\natexlab{b}}.
\newblock \doi{10.1103/PhysRevLett.128.120601}.

\bibitem[Martin et~al.(1973)Martin, Siggia, and Rose]{MSR_formalism}
P.~C. Martin, E.~D. Siggia, and H.~A. Rose.
\newblock Statistical dynamics of classical systems.
\newblock \emph{Phys. Rev. A}, 8:\penalty0 423--437, Jul 1973.
\newblock \doi{10.1103/PhysRevA.8.423}.

\bibitem[Megard et~al.(1987)Megard, Parisi, and Virasoo]{SpinGlassesBook}
M.~Megard, G.~Parisi, and M.~A. Virasoo.
\newblock \emph{Spin glass theory and beyond}, volume~9 of \emph{World
  Scientific lecture notes in physics}.
\newblock World Scientific, 1987.
\newblock ISBN 9971501155,9789971501150,9971501163,9789971501167.

\bibitem[Bunin(2016)]{bunin2016interaction}
Guy Bunin.
\newblock Interaction patterns and diversity in assembled ecological
  communities, 2016.
\newblock URL \url{https://arxiv.org/abs/1607.04734}.

\bibitem[Carlson(1972)]{Log_mean}
B.~C. Carlson.
\newblock The logarithmic mean.
\newblock \emph{The American Mathematical Monthly}, 79\penalty0 (6):\penalty0
  615--618, 1972.
\newblock ISSN 00029890, 19300972.
\newblock URL \url{http://www.jstor.org/stable/2317088}.

\bibitem[Matthews and Whittaker(2015)]{SAD_review_Matthews}
T.~J. Matthews and R.~J. Whittaker.
\newblock Review: On the species abundance distribution in applied ecology and
  biodiversity management.
\newblock \emph{Journal of Applied Ecology}, 52\penalty0 (2):\penalty0
  443--454, 2015.
\newblock \doi{https://doi.org/10.1111/1365-2664.12380}.

\bibitem[McGill et~al.(2007)McGill, Etienne, Gray, Alonso, Anderson, Benecha,
  Dornelas, Enquist, Green, He, Hurlbert, Magurran, Marquet, Maurer, Ostling,
  Soykan, Ugland, and White]{SAD_review_McGill}
B.~J. McGill, R.~S. Etienne, J.~S. Gray, D.~Alonso, M.~J. Anderson, H.~K.
  Benecha, M.~Dornelas, B.~J. Enquist, J.~L. Green, F.~He, A.~H. Hurlbert,
  A.~E. Magurran, P.~A. Marquet, B.~A. Maurer, A.~Ostling, C.~U. Soykan, K.~I.
  Ugland, and E.~P. White.
\newblock Species abundance distributions: moving beyond single prediction
  theories to integration within an ecological framework.
\newblock \emph{Ecology Letters}, 10\penalty0 (10):\penalty0 995--1015, 2007.
\newblock \doi{https://doi.org/10.1111/j.1461-0248.2007.01094.x}.

\bibitem[Yoshino et~al.(2008)Yoshino, Galla, and Tokita]{RADs_Yoshino}
Y.~Yoshino, T.~Galla, and K.~Tokita.
\newblock Rank abundance relations in evolutionary dynamics of random
  replicators.
\newblock \emph{Phys. Rev. E}, 78:\penalty0 031924, Sep 2008.
\newblock \doi{10.1103/PhysRevE.78.031924}.

\bibitem[Magurran(2011)]{magurran_2011}
Anne~E. Magurran.
\newblock \emph{Measuring biological diversity}.
\newblock Blackwell, 2011.

\bibitem[Kuczala and Sharpee(2016)]{Kuczala_block_variances}
A.~Kuczala and T.~O. Sharpee.
\newblock Eigenvalue spectra of large correlated random matrices.
\newblock \emph{Phys. Rev. E}, 94:\penalty0 050101, Nov 2016.
\newblock \doi{10.1103/PhysRevE.94.050101}.

\bibitem[Grilli et~al.(2015)Grilli, Barabás, and
  Allesina]{Grilli2015_metapopulaitons}
J~Grilli, G~Barabás, and S~Allesina.
\newblock Metapopulation persistence in random fragmented landscapes.
\newblock \emph{PLOS Computational Biology}, 11\penalty0 (5):\penalty0 1--13,
  05 2015.
\newblock \doi{10.1371/journal.pcbi.1004251}.

\bibitem[Hanski and Ovaskainen(2000)]{Hanski2000_metapopulations}
I~Hanski and O~Ovaskainen.
\newblock The metapopulation capacity of a fragmented landscape.
\newblock \emph{Nature}, 404\penalty0 (6779):\penalty0 755--758, Apr 2000.
\newblock ISSN 1476-4687.
\newblock \doi{10.1038/35008063}.

\bibitem[Johnson et~al.(2014)Johnson, Domínguez-García, Donetti, and
  A.]{Johnson2014_TrophicCoherence}
S.~Johnson, V.~Domínguez-García, L.~Donetti, and Muñoz.~M. A.
\newblock Trophic coherence determines food-web stability.
\newblock \emph{Proceedings of the National Academy of Sciences}, 111\penalty0
  (50):\penalty0 17923--17928, 2014.
\newblock \doi{10.1073/pnas.1409077111}.

\bibitem[Baiser et~al.(2013)Baiser, Whitaker, and
  Ellison]{Baiser2013_foundationSpecies}
B.~Baiser, N.~Whitaker, and A.~M. Ellison.
\newblock Modeling foundation species in food webs.
\newblock \emph{Ecosphere}, 4\penalty0 (12):\penalty0 art146, 2013.
\newblock \doi{https://doi.org/10.1890/ES13-00265.1}.

\bibitem[Hertz et~al.(2016)Hertz, Roudi, and Sollich]{Hertz_2016}
J.~A. Hertz, Y.~Roudi, and P.~Sollich.
\newblock Path integral methods for the dynamics of stochastic and disordered
  systems.
\newblock \emph{Journal of Physics A: Mathematical and Theoretical},
  50\penalty0 (3):\penalty0 033001, dec 2016.
\newblock \doi{10.1088/1751-8121/50/3/033001}.

\bibitem[De~Dominicis and Peliti(1978)]{De_Dominics_Peliti1978}
C.~De~Dominicis and L.~Peliti.
\newblock Field-theory renormalization and critical dynamics above ${T}_{c}$:
  Helium, antiferromagnets, and liquid-gas systems.
\newblock \emph{Phys. Rev. B}, 18:\penalty0 353--376, Jul 1978.
\newblock \doi{10.1103/PhysRevB.18.353}.

\bibitem[Janssen(1976)]{Janssen1976}
H.K. Janssen.
\newblock On a lagrangean for classical field dynamics and renormalization
  group calculations of dynamical critical properties.
\newblock \emph{Zeitschrift f{\"u}r Physik B Condensed Matter}, 23\penalty0
  (4):\penalty0 377--380, Dec 1976.
\newblock ISSN 1431-584X.
\newblock \doi{10.1007/BF01316547}.

\bibitem[Roy et~al.(2019)Roy, Biroli, Bunin, and Cammarota]{roy2019numerical}
F~Roy, G~Biroli, G~Bunin, and C~Cammarota.
\newblock Numerical implementation of dynamical mean field theory for
  disordered systems: application to the lotka{\textendash}volterra model of
  ecosystems.
\newblock \emph{Journal of Physics A: Mathematical and Theoretical},
  52\penalty0 (48):\penalty0 484001, nov 2019.
\newblock \doi{10.1088/1751-8121/ab1f32}.

\end{thebibliography}

\end{document}